\documentclass[preprint]{aastex62}
\usepackage{graphicx}
\usepackage{natbib}
\usepackage{amsmath}
\usepackage{bm}
\usepackage{gensymb}
\usepackage{float}
\usepackage{lineno}
\usepackage{aasref}
\usepackage{afterpage}
\usepackage{float}

\graphicspath{{Images/}}

\begin{document}

\title{A Survey of Cassini Images of Spokes in Saturn's Rings: \\ Unusual Spoke Types and Seasonal Trends}
\author{S. R. Callos}
\affil{Department of Physics, University of Oregon, Eugene OR 97403}
 \author{M.M. Hedman}
 \affil{Department of Physics, University of Idaho, Moscow ID 83844}
 \author{D.P. Hamilton}
 \affil{Astronomy Department, University of Maryland, College Park MD 20742}

\begin{abstract}
Spokes are localized clouds of fine particles that appear over the outer part of Saturn's B ring. Over the course of the Cassini Mission, the Imaging Science Subsystem (ISS) obtained over 20,000 images of the outer B ring, providing the most comprehensive data set for quantifying spoke properties currently available. Consistent with prior work, we find that spokes typically appear as dark features when the lit side of the rings are viewed at low phase angles, and as bright features when the rings are viewed at high phase angles or the dark side of the rings are observed. However, we also find examples of spokes on the dark side of the rings that transition between being brighter and darker than the background ring as they move around the planet. Most interestingly, we also identify spokes that appear to be darker than the background ring near their center and brighter than the background ring near their edges. These ``mixed spokes'' indicate that the particle size distribution can vary spatially within a spoke. In addition, we document seasonal variations in the overall spoke activity over the course of the Cassini mission using statistics derived from lit-side imaging sequences.  These statistics demonstrate that while spokes can be detected over a wide range of solar elevation angles, spoke activity increases dramatically when the Sun is within 10$^\circ$ of the ring plane.
\end{abstract}
 
\section{Introduction}

Spokes are forward-scattering clouds of micron-sized particles found above Saturn’s B-ring \citep{Collins80, Porco82, D'Aversa10} that can appear as light or dark patches depending on the lighting geometry \citep{Smith81, Grun83}. They were first officially  seen in images obtained by Voyager 1 in 1980 and Voyager 2 in 1981 \citep{Collins80, Smith81}. However, there are some indications they were seen earlier than this. Stephan O’meara reported seeing them in 1976, when the ring plane opening angle was approximately 15$\degree$. There are also reports of ‘clouds,’ ‘spots,’ and other peculiarities appearing within the rings of Saturn going as far back as the 1700s \citep{Alexander62, Robinson80}. The Voyager data showed that spoke activity was correlated with asymmetries in Saturn's magnetosphere \citep{Porco82}, while more recent observations revealed that the prevalence of spokes varies with  Saturn’s seasons, with them occurring most prominently around Saturn’s equinoxes \citep{McGhee05, Horanyi09, Simon23}. 

Multiple explanations have been proposed for how these features form over the B ring. The seasonality of the spokes' appearance and their correlation with asymmetries in Saturn's magnetosphere \citep{Porco82, Mitchell13} strongly suggest that non-gravitational processes are involved. These features could be triggered by electric storms on Saturn \citep{Jones06} or meteoroid impacts \citep{Goertz83} and the dusty material released by these events could be dispersed by a plasma cloud, electron beams, plasma waves or avalanches of small dust particles \citep{Hill82, Goertz83, Tagger91, Grun94,  Yaroshenko08, Hirata22, Farmer05, Morfill05, Hamilton06}. However, thus far none of these theoretical ideas have been fully tested against the full span of the Cassini observations. While Cassini obtained over 20,000 images of spokes over the course of the mission, published studies of those data have focused primarily on only a few observations from time periods when the spokes first became obvious \citep{Mitchell06} and when the spokes were strongly active \citep{Mitchell13}.  This study provides a more comprehensive look of the available spoke observations that should help facilitate future investigations of their properties and origins.

Section~\ref{catalog} describes how we cataloged many of the spoke observations obtained by the Cassini cameras and how the data from each image were transformed into standardized maps of ring brightness versus radius and longitude that facilitate comparisons among different observations.  Section~\ref{survey} provides a high-level qualitative summary of these data, including which observations contained visible spokes and whether those spokes appeared as dark or bright features. This survey also revealed some unusual and previously unreported types of spokes, including spokes that could appear as either bright or dark depending on where they were observed, and others that have both bright and dark components. Next, Section~\ref{quantify} examines how spoke activity varies over time. We identify a small number of faint spokes visible late in the Cassini mission. In addition, we develop some simple metrics for average spoke activity that we apply to a sub-set of the lit-side spoke images. These data provide a clearer picture of how spoke activity changes with Saturn's seasons. Finally, Section~\ref{summary} summarizes the findings of this preliminary study.

\section{Catalog of spoke observations}
\label{catalog}

We searched for images of spokes obtained by the Imaging Science Subsystem (ISS) onboard the Cassini Spacecraft \citep{Porco04} using the Outer Planets Unified Search tool (OPUS, https://pds-rings.seti.org/search/) provided by the Planetary Data System's Ring-Moon Systems node. First, we searched for all images that were part of observations whose names contained ‘SPOKE’ or ‘SPK’ because each of these observations contained multiple images obtained over a short time period that were designed to observe spokes as they moved across the rings, and so provide the most useful information about overall spoke intensity. These observations provided adequate coverage for  2004-2009 and 2013-2014. However, there are few lit side spoke images taken between Cassini orbits (or `Revs') 117 and 199 (corresponding to years 2010-2012) due to the low inclination of the spacecraft's orbit during this time. Since this includes the time period when spoke activity declined after equinox, we used OPUS to search for any images of the rings that might contain spokes. Specifically, we searched for images with observed north-based incidence angle between 72 and 89 degrees (to see images taken from 2010 to 2012), and observed ring radii between 92,000 and 117,000 kilometers. We selected images with a maximum emission angle of 85$^\circ$ and a maximum observed resolution of 100 km/pixel and included those observations in our catalog. Table~\ref{tab:allobs} summarizes the basic properties of the observations considered in this study, including the relevant observation and image names, some important aspects of the viewing and lighting geometry, and what types of spokes are visible in the images (see below).

Each image was calibrated using the standard {CISSCAL 4.0beta} pipeline available on the Planetary Data System, which removes instrumental backgrounds, applies flat fields, and converts the brightness data to $I/F$, a standardized measure of reflectance \citep{West10, Knowles20}. Each calibrated image was also geometrically navigated by first using the appropriate SPICE kernels \citep{Acton96} to roughly estimate the location and orientation of the rings in the camera's field of view. For most images, the opaque B ring filled so much of the field of view that stars could not be used to refine the image pointing. We therefore used the nominal geometry to compute estimates of the ringplane radius and inertial longitude for each image pixel, along with how these parameters varied with pixel location. We then used this information to divide the image into 10 regions corresponding to different ranges of inertial longitudes, and computed the average brightness as a function of radius for each of these regions. Each profile was then cross-correlated with a reference ring brightness profile derived from a single reference image (W1597993945 for lit-side images and W1545330616 for dark-side images). The location of the peak cross-correlation was then used to estimate the radial shift needed to align each of the 10 profiles, which was in turn translated into estimates of the required pixel offsets using linear regression based on the average dependence of how the radius depended on pixel coordinates in each profile. This algorithm was sufficiently automatic and efficient to process nearly all of the required images, provided we considered the brightness profiles only in the region between 90,000 and 115,000 km. For the approximately 300 images where the algorithm was unable to sufficiently repoint the images,\footnote{Most of theses images belong to the observation sequences ISS\_00BRI\_SPKMOVPER001\_PRIME, ISS\_00BRI\_SPKFORM002\_PRIME, ISS\_058RI\_SPKFORM001\_PRIME, ISS\_116RI\_HPSPOKE001\_VIMS, ISS\_168RI\_SPOKEMOV001\_PRIME, and ISS\_200RI\_SPOKEMOV010\_PRIME}  manual repointing was done using moons and background stars as reference points. 

After each image was navigated, we used the relevant geometric information to compute the radius and inertial longitude for each pixel in the image. The data were then re-projected onto regular grids of radii and inertial longitude to produce standardized maps (a total of 6,177 maps) of the B-ring's brightness versus radius and inertial longitude relative the ring's ascending node on the J2000 equatorial plane. These maps will be archived by the Planetary Data System \citep{SpokeDB} and should facilitate identifying and quantifying spokes because each row of the map corresponds to one radius in the ring. The longitudinal brightness variations due to the spokes can therefore be isolated from the radial brightness variations from the background B ring by subtracting either the median brightness or a low-order polynomial fit from each row of the image.

\begin{table}[!ht]
\caption{Overview of Cassini Spoke Observations}
\label{tab:allobs}
\hspace{-0.5in}    \resizebox{4in}{!}{\begin{tabular}{|l|l|l|l|l|l|l|}
    \hline
        Observation Sequence & Start/End Image & Type$^a$ & Solar El. & Emission  & Phase & Num.  \\ 
                & & & Angle $^b$ ($^\circ$) & Angle$^c$ ($^\circ$) & Angle$^d$ ($^\circ$) & Images \\ \hline
        ISS\_00ARI\_SPKMOVPER001\_PRIME & N1479201492/N1479254052 & X & -23.4 & 102.77 & 84.6 & 69 \\ \hline
        ISS\_00BRI\_SPKMOVPER002\_PRIME & W1479724035/W1479962645 & X & -23.34 & 101.9 & 76.1 & 91 \\ \hline
        ISS\_00BRI\_SPKFORM001\_PRIME & N1481263615/N1481270245 & X & -23.19 & 97.31 & 50.8 & 426 \\ \hline
        IOSIC\_006RI\_SUBMU04HP001\_SI & W1492259467/W1492262463 & X & -22 & 86.32 & 135.8 & 6 \\ \hline
        ISS\_007RI\_SPKLRSLPA001\_PRIME & N1493107982/W1493127183 & X & -21.9 & 103.58 & 57.9 & 59 \\ \hline
        ISS\_007RI\_SPKLRSLPA002\_PRIME & W1493187183/W1493206383 & X & -21.88 & 105.45 & 50.7 & 59 \\ \hline
        IOSIC\_008RI\_VERTULHP003\_SI & W1495413804/W1495416684 & X & -21.63 & 79.36 & 134 & 7 \\ \hline
        ISS\_011RI\_SPKLRSLPA001\_PRIME & N1499479804/N1499496561 & X & -21.14 & 105.53 & 59.4 & 14 \\ \hline
        ISS\_011RI\_SPKLRSLPB001\_PRIME & N1499499604/N1499516361 & X & -21.14 & 105.74 & 59.1 & 14 \\ \hline
        ISS\_011RI\_SPKLRSLPA002\_PRIME & N1499562605/N1499579362 & X & -21.13 & 106.56 & 55.2 & 14 \\ \hline
        ISS\_011RI\_SPKLRSLPB002\_PRIME & N1499582405/N1499599162 & X & -21.13 & 106.84 & 54.8 & 14 \\ \hline
        ISS\_011RI\_SPKLRSLPA003\_PRIME & N1499649005/N1499663514 & X & -21.12 & 107.53 & 49.8 & 12 \\ \hline
        ISS\_011RI\_SPKLRSLPB003\_PRIME & N1499667005/N1499679534 & X & -21.12 & 107.69 & 48.6 & 12 \\ \hline
        IOSIC\_012RI\_TEMPU05HP001\_SI & W1501734543/W1501735983 & X & -20.86 & 82.94 & 130 & 2 \\ \hline
        IOSIC\_014RI\_SUBMU12HP001\_SI & W1504634392/W1504636012 & V & -20.49 & 73.63 & 145.2 & 3 \\ \hline
        ISS\_035RI\_SPKMRHPDF001\_PRIME & W1545324916/W1545342016 & V & -14.53 & 43.65 & 140.7 & 17 \\ \hline
        ISS\_035RI\_SPKMRHPDF002\_PRIME & W1545407757/W1545423717 & V & -14.52 & 47.07 & 147.3 & 15 \\ \hline
        ISS\_036RI\_SPKDKSHAD001\_PRIME & W1546549044/W1546557397 & V & -14.34 & 36.96 & 105.3 & 48 \\ \hline
        ISS\_036RI\_SPKMRHPDF001\_PRIME & W1546786192/W1546805680 & V & -14.28 & 42.38 & 142 & 26 \\ \hline
        ISS\_036RI\_SPKMRHPDF002\_PRIME & W1546845422/W1546915219 & V & -14.28 & 43.83 & 144.5 & 92 \\ \hline
        ISS\_037RI\_SPKMRHPDF002\_PRIME & W1546958833/W1547006353 & V & -14.27 & 49.21 & 152.3 & 45 \\ \hline
        ISS\_058RI\_SPKFORM001\_PRIME & N1581466518/N1581476999 & V & -8.41 & 82.06 & 16.3 & 912 \\ \hline
        ISS\_063RB\_SPOKE001\_VIMS & W1585810269/W1585816103 & X & -7.64 & 140.92 & 50.3 & 12 \\ \hline
        ISS\_063RI\_SPKFORMLF001\_PRIME & W1585896379/W1585916540 & D & -7.63 & 104.41 & 17.1 & 7 \\ \hline
        ISS\_066RI\_SPKHRLPLF001\_PRIME & W1588324357/W1588333207 & D & -7.2 & 120.61 & 32.2 & 2 \\ \hline
        ISS\_066RI\_SPKSTFORM001\_PRIME & W1588432867/W1588433823 & D & -7.18 & 93.68 & 11.3 & 4 \\ \hline
        ISS\_067RI\_SPKFORMLF001\_PRIME & W1589180203/W1589197843 & D & -7.04 & 109.73 & 22.7 & 7 \\ \hline
        ISS\_071RI\_SPOKEMOV001\_PRIME & W1591782176/W1591793456 & D & -6.58 & 108.35 & 19.1 & 17 \\ \hline
        ISS\_071RI\_SPOKEMOV001\_PRIME & W1591782176/W1591793456 & D & -6.58 & 108.35 & 19.1 & 17 \\ \hline
        ISS\_072RI\_SPKHRLPDF001\_PRIME & W1592114050/W1592159350 & V & -6.52 & 64.23 & 32.5 & 76 \\ \hline
        ISS\_074RI\_SPKLFMOV001\_PRIME & W1593676089/W1593708713 & D & -6.25 & 97.99 & 14.4 & 17 \\ \hline
        ISS\_078RI\_SPKFORM001\_PRIME & W1595992406/W1596027399 & D & -5.83 & 127.76 & 36.8 & 68 \\ \hline
        ISS\_081RI\_SPKMVLFLP001\_PRIME & W1597971520/W1598009545 & D & -5.48 & 109.19 & 23.1 & 40 \\ \hline
        ISS\_085RI\_SPKMVLFLP001\_PRIME & W1600545658/W1600581086 & D & -5.01 & 104.39 & 21.5 & 35 \\ \hline
        ISS\_086RI\_SPKMVLFLP002\_PRIME & W1601160662/W1601196432 & D & -4.91 & 108.89 & 24.6 & 36 \\ \hline
        ISS\_088RI\_SPKMVLFLP001\_PRIME & W1602457751/W1602500234 & D & -4.68 & 103.6 & 21.9 & 50 \\ \hline
        ISS\_094RI\_SPKMVLFLP001\_PRIME & W1606343679/W1606379775 & D & -3.98 & 122.86 & 38.8 & 49 \\ \hline
        ISS\_096RI\_SPOKE105\_VIMS & W1607513021/W1607513903 & V & -3.77 & 45.44 & 111 & 7 \\ \hline
        ISS\_096RI\_SPKFMLFLP001\_PRIME & W1607710973/W1607725973 & D & -3.74 & 121.18 & 68.6 & 5 \\ \hline
        ISS\_102RI\_SPKTRKLF001\_PRIME & W1612298782/W1612322722 & B & -2.92 & 117.59 & 139.4 & 34 \\ \hline
        ISS\_102RI\_SPKFMLFLP001\_PRIME & W1612545569/W1612574369 & D & -2.87 & 118.52 & 40.5 & 9 \\ \hline
        ISS\_108RI\_SPKMVLFLP001\_PRIME & W1618050603/W1618070583 & M & -1.89 & 111.55 & 39.9 & 19 \\ \hline
        ISS\_109RI\_SPKFMLFHP001\_PRIME & W1619084400/W1619098580 & B & -1.7 & 144.46 & 123.5 & 5 \\ \hline
        ISS\_109RI\_LPSPOKE001\_VIMS & W1619722638/W1619755938 & V & -1.59 & 36.49 & 60.3 & 37 \\ \hline
        ISS\_110RI\_SPKFMDFHP001\_PRIME & W1620033982/W1620044782 & V & -1.53 & 50.89 & 112.2 & 4 \\ \hline
        ISS\_113RI\_SPKMVDFHP001\_PRIME & W1624063786/W1624260707 & V & -0.82 & 56.53 & 113.2 & 46 \\ \hline
        ISS\_115RI\_SPKMVDFHP001\_PRIME & W1626751925/W1626773525 & V & -0.34 & 66.91 & 116.4 & 37 \\ \hline
        ISS\_115RI\_SPKMVDFHP002\_PRIME & W1626811325/W1626856755 & V & -0.33 & 69.2 & 119.5 & 60 \\ \hline
        ISS\_115RI\_SPKMVDFHP003\_PRIME & W1626901926/W1626938001 & V & -0.31 & 73.13 & 124.7 & 38 \\ \hline
        ISS\_115RI\_SPKMVDFHP004\_PRIME & W1626984126/W1627021746 & V & -0.3 & 76.8 & 135.9 & 39 \\ \hline
        ISS\_116RI\_SPKMVDFHP001\_PRIME & W1628044034/W1628093687 & V & -0.11 & 71.9 & 114.3 & 69 \\ \hline
        ISS\_116RI\_SPKMVDFHP002\_PRIME & W1628132834/W1628178823 & V & -0.09 & 74.92 & 124.2 & 60 \\ \hline
        ISS\_116RI\_SPKMVDFHP003\_PRIME & W1628224335/W1628261267 & V & -0.07 & 76.5 & 122.5 & 57 \\ \hline
        ISS\_116RI\_SPKMVDFHP004\_PRIME & W1628308036/W1628350133 & V & -0.06 & 79.13 & 126.8 & 55 \\ \hline
        ISS\_116RI\_HPSPOKE001\_VIMS & W1629233206/W1629289474 & X & 0.11 & 77.87 & 114.4 & 7 \\ \hline
        ISS\_117RI\_SPKMVLFHP001\_PRIME & W1630344949/W1630392679 & B & 0.31 & 78.35 & 78.4 & 74 \\ \hline
        ISS\_117RI\_SPKMVLFHP003\_PRIME & W1630444550/W1630507770 & B & 0.32 & 78.86 & 84.5 & 110 \\ \hline
        ISS\_117RI\_SPKMVLFHP004\_PRIME & W1630543551/W1630594281 & B & 0.34 & 79.33 & 89.2 & 90 \\ \hline
        ISS\_117RI\_SPKMVLFHP005\_PRIME & W1630629951/W1630678551 & B & 0.35 & 79.72 & 92.5 & 82 \\ \hline
        ISS\_117RI\_SPKMVLFHP006\_PRIME & W1630723552/W1630788645 & B & 0.37 & 80.21 & 96.6 & 119 \\ \hline
        ISS\_117RI\_SPKMVLFHP007\_PRIME & W1630826153/W1630871681 & B & 0.39 & 80.67 & 99.6 & 85 \\ \hline
        \end{tabular}}
\hspace{-0.5in}   \resizebox{4in}{!}{\begin{tabular}{|l|l|l|l|l|l|l|}
    \hline
        Observation Sequence & Start/End Image & Type$^a$ & Solar El. & Emission  & Phase & Number of  \\ 
        & & & Angle $^b$ ($^\circ$) & Angle$^c$ ($^\circ$) & Angle$^d$ ($^\circ$) & Images \\ \hline
             ISS\_117RI\_SPKMVLFHP008\_PRIME & N1630912553/N1630959462 & B & 0.41 & 81.03 & 101.6 & 62 \\ \hline
        ISS\_117RI\_SPKMVLFHP009\_PRIME & N1631002166/N1631027290 & B & 0.42 & 81.27 & 103.1 & 23 \\ \hline
        ISS\_117RI\_SPKMVLFHP010\_PRIME & N1631083166/N1631135042 & B & 0.43 & 81.61 & 105.4 & 45 \\ \hline
        ISS\_118RI\_SPKMVLFHP001\_PRIME & N1631177667/N1631220542 & B & 0.45 & 82.03 & 108 & 36 \\ \hline
        ISS\_118RI\_SPKMVLFHP002\_PRIME & N1631341856/N1631361096 & B & 0.48 & 82.8 & 112.6 & 27 \\ \hline
        ISS\_118RI\_SPKMVLFHP003\_PRIME & N1631411069/N1631455974 & B & 0.49 & 83.18 & 114.9 & 36 \\ \hline
        ISS\_118RI\_SPKMVLFHP004\_PRIME & N1631497469/N1631532947 & B & 0.51 & 83.7 & 117.8 & 28 \\ \hline
        ISS\_124RI\_SPKMVDFHP001\_PRIME & N1641477428/N1641522428 & V & 2.27 & 104.34 & 113.4 & 76 \\ \hline
        IOSIC\_124RI\_EQLBN002\_SI & W1641929271/W1641933584 & B & 2.35 & 80.7 & 56.9 & 5 \\ \hline
        IOSIC\_132RI\_EQLBN001\_SI & W1653990287/W1653991810 & B & 4.44 & 80.61 & 115.1 & 4 \\ \hline
        IOSIC\_132RI\_EQLBN003\_SI & W1654130748/W1654133020 & B & 4.47 & 79.32 & 133.7 & 5 \\ \hline
        IOSIC\_132RI\_EQLBN004\_SI & W1654136515/W1654140415 & B & 4.47 & 79.32 & 133.7 & 4 \\ \hline
        ISS\_134RI\_SPKMVDFHP001\_PRIME & W1656595136/W1656635176 & V & 4.89 & 102.81 & 107.6 & 66 \\ \hline
        ISS\_134RI\_SPKMVDFHP002\_PRIME & W1656702236/W1656738426 & V & 4.91 & 104.26 & 113.9 & 56 \\ \hline
        ISS\_134RI\_SPKMVDFHP003\_PRIME & W1656799778/W1656857318 & V & 4.93 & 106.25 & 124.3 & 15 \\ \hline
        IOSIC\_134RI\_P50L30S15002\_SI & W1657016906/W1657037174 & D & 4.96 & 72.77 & 19.6 & 6 \\ \hline
        IOSIC\_134RI\_P50L30S15002\_SI & W1657038894/W1657048173 & D & 4.97 & 80.1 & 19 & 6 \\ \hline
        IOSIC\_137RI\_EQLBS001\_SI & W1662110415/W1662112806 & X & 5.83 & 94.73 & 150.67 & 6 \\ \hline
        IOSIC\_137RI\_EQLBN001\_SI & W1662216342/W1662218563 & X & 5.85 & 87.18 & 9.87 & 6 \\ \hline
        IOSIC\_137RI\_EQLBN002\_SI & W1662228582/W1662230803 & X & 5.85 & 87.92 & 41 & 6 \\ \hline
        ISS\_168RI\_SPOKEMOV001\_PRIME & W1719216742/W1719283410 & V & 14.81 & 109.94 & 98 & 64 \\ \hline
        ISS\_168RB\_BMOVIE001\_PRIME & W1719710671/W1719746078 & D & 14.88 & 82.2 & 15.9 & 6 \\ \hline
        ISS\_172RI\_SPOKEMOV001\_PRIME & W1726689810/W1726727886 & V & 15.85 & 119.4 & 93.1 & 58 \\ \hline
        ISS\_172RI\_SPOKEMOV002\_PRIME & W1726771710/W1726808835 & V & 15.87 & 120.63 & 104.3 & 56 \\ \hline
        IOSIC\_172RI\_NP20L30S001\_SI & W1727250659/W1727255245 & X & 15.93 & 67.11 & 27 & 3 \\ \hline
        ISS\_173RI\_SPOKEMOV001\_PRIME & W1728245860/W1728279160 & V & 16.07 & 124.47 & 93 & 46 \\ \hline
        ISS\_173RI\_SPOKEMOV002\_PRIME & W1728607763/W1728626213 & V & 16.11 & 127.94 & 104.1 & 151 \\ \hline
        ISS\_173RI\_SPOKEMOV003\_PRIME & W1728757643/W1728822864 & V & 16.13 & 129.14 & 113 & 174 \\ \hline
        IOSIC\_173RI\_NP50L30N001\_SI & W1729270212/W1729274798 & X & 16.2 & 48.19 & 72.1 & 3 \\ \hline
        ISS\_174RI\_SPOKEMOV001\_PRIME & W1730573575/W1730692915 & V & 16.38 & 127.43 & 99 & 118 \\ \hline
        ISS\_174RI\_SPOKEMOV002\_PRIME & W1730746588/W1730859748 & V & 16.4 & 128.82 & 107.9 & 93 \\ \hline
        IOSIC\_180RI\_NP50L70S001\_SI & W1738444466/W1738536506 & X & 16.41 & 39.38 & 38.4 & 4 \\ \hline
        ISS\_174RI\_SPOKEMOV004\_PRIME & W1731024177/W1731060057 & V & 16.44 & 126.65 & 129.8 & 47 \\ \hline
        IOSIC\_176RI\_NP50L70004\_SI & W1733851636/W1733898480 & X & 16.82 & 44.56 & 48.01 & 8 \\ \hline
        ISS\_177RI\_SPOKEMOV001\_PRIME & W1734640040/ W1734683888 & V & 16.92 & 138.21 & 97.1 & 57 \\ \hline
        ISS\_178RI\_SPOKEMOV001\_PRIME & W1735820948/W1735861586 & V & 17.07 & 139.28 & 103.1 & 27 \\ \hline
        ISS\_179RI\_SPOKEMOV001\_PRIME & W1737000455/W1737047057 & V & 17.23 & 140.29 & 108.1 & 28 \\ \hline
        IOSIC\_179RI\_NP20L30S001\_SI & W1737229803/W1737354769 & X & 17.26 & 67.1 & 117.3 & 4 \\ \hline
        ISS\_180RI\_SPOKEMOV001\_PRIME & W1738087703/W1738104272 & V & 17.37 & 137.91 & 97.3 & 64 \\ \hline
        ISS\_180RI\_SPOKEMOV002\_PRIME & W1738174663/W1738201319 & V & 17.38 & 140.51 & 111.7 & 33 \\ \hline
        ISS\_181RI\_SPOKEMOV001\_PRIME & W1739296670/ W1739313067 & V & 17.52 & 140.23 & 106.4 & 20 \\ \hline
        IOSIC\_181RI\_NP20L70S001\_SI & W1739633073/W1739656233 & X & 17.57 & 58.27 & 14.2 & 8 \\ \hline
        ISS\_183RI\_SPOKEMOV001\_PRIME & W1741302783/W1741345070 & V & 17.78 & 144.77 & 104.4 & 50 \\ \hline
        IOSIC\_183RI\_NP20L30S001\_SI & W1741703126/W1741729336 & D & 17.83 & 61.46 & 10.8 & 8 \\ \hline
        ISS\_184RI\_SPOKEMOV002\_PRIME & W1742314089/W1742329623 & V & 17.91 & 143.25 & 100.2 & 19 \\ \hline
        ISS\_198RI\_SPOKEMOV004\_PRIME & W1760871907/W1760917141 & V & 20.12 & 136.87 & 120 & 64 \\ \hline
        ISS\_198RI\_SPOKEMOV005\_PRIME & W1761058908/W1761141862 & V & 20.14 & 129.88 & 139.1 & 60 \\ \hline
        ISS\_199RI\_SPOKEMOV002\_PRIME & W1762866539/W1762968219 & B & 20.34 & 77.32 & 144 & 122 \\ \hline
        ISS\_199RI\_SPOKEMOV003\_PRIME & W1763013921/W1763139321 & B & 20.35 & 73.18 & 138.4 & 151 \\ \hline
        ISS\_199RI\_SPOKEMOV004\_PRIME & W1763215522/W1763316294 & B & 20.37 & 67.35 & 131.1 & 119 \\ \hline
        ISS\_199RI\_SPOKEMOV006\_PRIME & W1763480768/W1763637247 & B & 20.4 & 59.28 & 120.1 & 168 \\ \hline
        ISS\_200RI\_SPOKEMOV001\_PRIME & W1766030604/W1766079224 & B & 20.68 & 68.34 & 133.4 & 56 \\ \hline
        ISS\_200RI\_SPOKEMOV004\_PRIME & W1766474355/W1766591421 & B & 20.73 & 47.65 & 104.5 & 109 \\ \hline
        ISS\_200RI\_SPOKEMOV011\_PRIME & W1768356955/W1768399835 & B & 20.93 & 75.17 & 141.8 & 65 \\ \hline
        ISS\_201RI\_SPOKEMOV011\_PRIME & W1771092912/W1771136472 & X & 21.21 & 61.77 & 124 & 67 \\ \hline
        ISS\_201RI\_SPOKEMOV013\_PRIME & W1771266373/W1771310533 & X & 21.23 & 55.2 & 114.4 & 61 \\ \hline
        ISS\_201RI\_SPOKEMOV015\_PRIME & W1771441574/W1771491074 & X & 21.55 & 49.58 & 105.1 & 67 \\ \hline
        IOSIC\_206RI\_SP160L7001\_SI & W1784314522/W1784416428 & V & 22.51 & 121.86 & 137.5 & 2 \\ \hline
        IOSIC\_207RC\_COMPLLRES001\_SI & W1786529457/W1786552082 & D & 22.7 & 73.17 & 16.68 & 6 \\ \hline
        ISS\_208SA\_EUVFUV002\_UVIS & W1788965701/W1788965735 & X & 22.92 & 50.87 & 53.87 & 2 \\ \hline
        ISS\_269RI\_HIPHASEFB001\_PIE & W1870770709/W1870770784 & V & 26.73 & 116.6 & 174.9 & 5 \\ \hline
    \end{tabular}}
    
\footnotesize    $^a$ Type of spokes visible in the images. X = No spokes visible, D = Dark spokes, B = Bright Spokes, M = Mixed Spokes, V = Variable Spokes (all spokes on dark side of the rings).
    
    $^b$ Solar Elevation Angle relative to the ring plane, which is negative if the Sun is on the south side of the rings and positive if the Sun is on the north side of the rings
    
    $^c$ Emission Angle, which is the angle between the line of sight to the spacecraft and the {ring's northward} surface normal. Emission angles below 90$^\circ$ mean the spacecraft views the north side of the rings and emission angles above 90$^\circ$ mean the spacecraft views the south side of the rings.
    
    $^d$ Phase Angle, which is the angle between the incident sunlight and the emitted reflected light from the rings. 
    
\end{table}

\section{Qualitative summary of spoke properties}
\label{survey}

\begin{figure*}
\resizebox{\textwidth}{!}{\includegraphics{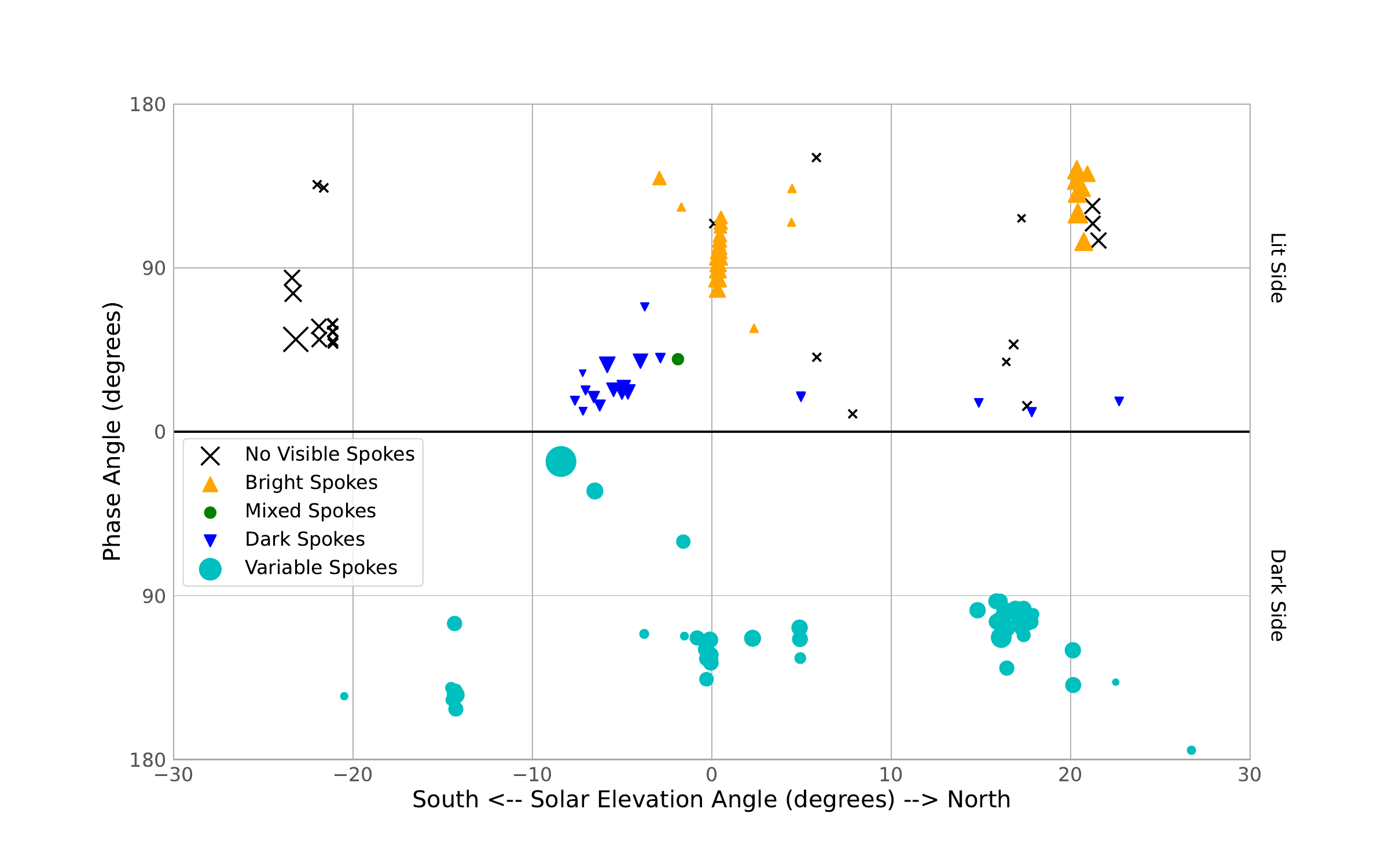}}
\caption{Timeline of spoke observations, showing the observed phase angle as a function of time (expressed in terms of the solar elevation angle) for images obtained on both the lit and dark sides of the rings. Each point corresponds to a particular Cassini ISS observation of the B ring, with the size of the symbol indicating the number of images in the observation, and the symbol shape/color indicating what type of spokes are present in these observations. The lit side observations can be reasonably well categorized as containing bright, dark or mixed spokes, while the appearance of the spokes seen on the dark side are more difficult to categorize (see Section~\ref{variables}).}
\label{timeline}
\end{figure*}

Figure~\ref{timeline} provides a graphical timeline of the Cassini spoke observations listed in Table~\ref{tab:allobs}. Note that the horizontal axis on this timeline is solar elevation angle above the ring plane, which is negative when the sun is on the south side of the rings and positive when the sun is on the north side of the rings. The  solar elevation angle is therefore a monotonic (but nonlinear) function of time for the duration of the Cassini mission, with early Cassini observations falling on the left side of the plot and observations near the end of the mission appearing towards the right side of the plot. Plotting spoke properties versus solar elevation angle rather than time is useful because solar illumination is expected to play an important role in the spoke's visibility  \citep{Nitter98, McGhee05, Horanyi09, Simon23}. This plot also illustrates the observed phase angle of the rings during each observation and whether the observed side of the rings was facing the sun or not.

The colors and shape of the symbols indicate whether spokes were visible in the observations, and what types of spokes were seen in the images and/or maps. For this survey, we found it useful to categorize the observed spokes into four types: dark, bright, mixed and variable. Note that these categories are orthogonal to the three morphological types (extended, narrow and filamentary) created by \citet{Grun83}, and are primarily useful for characterizing the overall appearance of the spokes in a particular sequence. Prior analyses of Voyager and Cassini observations already showed that spokes can appear as either dark features or bright features, depending on the observation and illumination geometry \citep{Smith81, Mitchell13}. However, for reasons that will become clear shortly, we only designate spokes as dark or bright when they appear on the lit side of the rings.

\begin{figure*}[tbp]
    \centering
    \includegraphics[width=6.5in]{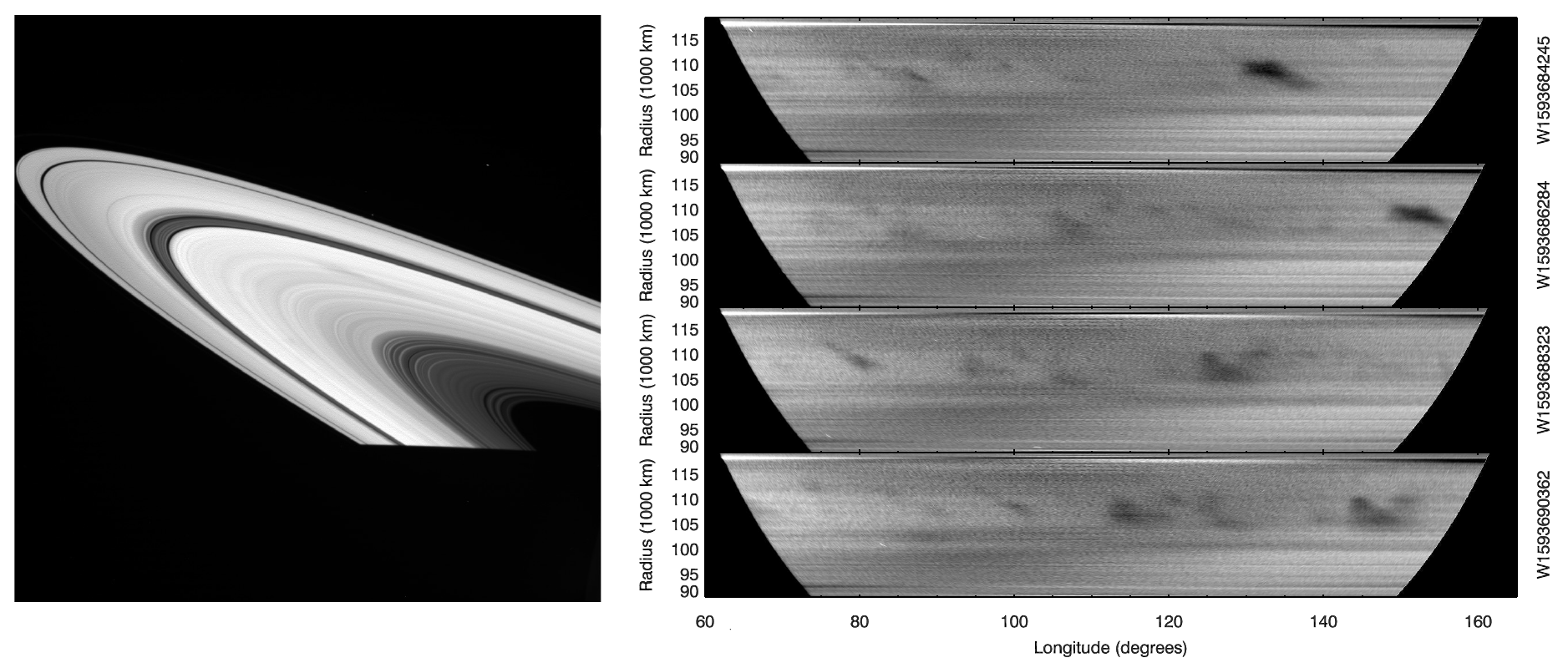}
    \caption{Examples of dark spokes seen on the lit side of the rings. The left panel shows a representative image (W1593684245) from observation sequence ISS\_074RI\_SPKLFMOV001\_PRIME, which was obtained {on July 2, 2008} at a phase angle of 14$^\circ$ when the solar elevation angle was -6.25$^\circ$. Note the dark spoke visible as a long diagonal slash on the upper side of the B ring. The images on the right are maps of the B ring's brightness versus radius and inertial longitude derived from several images in this same observation sequence. The median brightness at each radius in the illuminated part of the ring has been removed from each of these maps so that spoke signals can be more easily seen. In this case all the visible spokes appear as dark patches that move from left to right over time.  The spoke visible in the image on the left corresponds to the dark feature at 107,000 km and 140$^\circ$ longitude in the top map.}
    \label{fig:074}
\end{figure*}

\begin{figure*}[btp]
    \centering
    \includegraphics[width=6.5in]{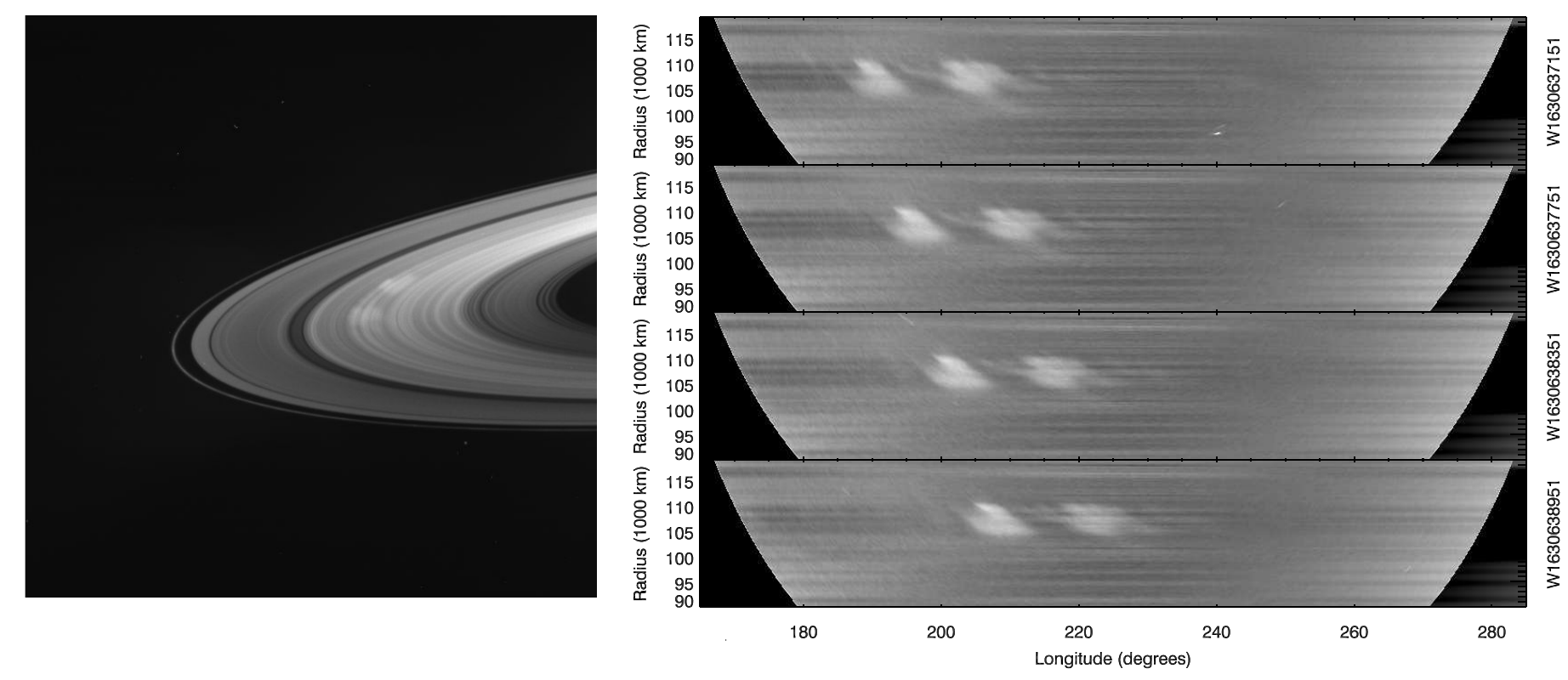}
    \caption{Examples of bright spokes seen on the lit side of the rings. The left panel shows a representative image (W1630638951) from observation sequence ISS\_117RI\_SPKMVLFHP005\_PRIME, which was obtained {September 3, 2009} at a phase angle of 92.5$^\circ$ when the Solar Elevation angle was 0.34$^\circ$. Note the bright spokes visible on the upper part of the B ring. The images on the right are maps of the B ring's brightness versus radius and inertial longitude derived from several images in this same observation sequence. A linear trend has been removed at each radius from each of these maps so that spoke signals can be more easily seen.}
    \label{fig:117_005}
\end{figure*}

\subsection{Dark and bright spokes on the lit side of the rings}

Figures \ref{fig:074} and \ref{fig:117_005} show sample images and maps derived from observations where the spokes can clearly be identified as either dark (Figure~\ref{fig:074}) or bright (Figure~\ref{fig:117_005}).  For display purposes, these maps have been filtered by subtracting the average observed brightness at each radius, which removes background brightness variations in the B ring, making the spokes easier to see.

Consistent with prior work, we find that dark spokes occur when the lit side of the rings is viewed at lower phase angles and bright spokes are found when the lit side of the rings is viewed at higher phase angles. As can be seen in Figure~\ref{timeline}, the transition between these two types of spokes falls at phase angles somewhere between 45$^\circ$ and 90$^\circ$.  This transition arises because at low phase angles the spoke particles prevent some fraction of the light from the background B ring from reaching the observer, while at high phase angles the spoke particles themselves can effectively scatter sunlight into the camera. This situation arises naturally if the spoke particles are roughly micron-sized \citep{D'Aversa10}, which makes them both relatively efficient at scattering visible light and strongly forward-scattering. Note that the background B ring is also darker at higher phase angles, so the reduction in ring signal by the intervening spoke material also becomes less important at higher phase angles. Similarly, the lit side of the B ring becomes substantially darker when the solar elevation angle gets close to zero, potentially allowing spokes to appear bright at lower phase angles.

\subsection{Variable Spokes on the unlit side of the rings}
\label{variables}

Spokes observed on the unlit side of the rings often appear as bright features, which is reasonable since most observations of the unlit side of the rings are at high phase angles, where spoke particles should be efficient scatterers. However, our survey of the Cassini images revealed that the appearance of spokes on the unlit side of the rings can depend on where the spokes are observed.

\begin{figure*}[p]
    \centering
    \includegraphics[width=7in]{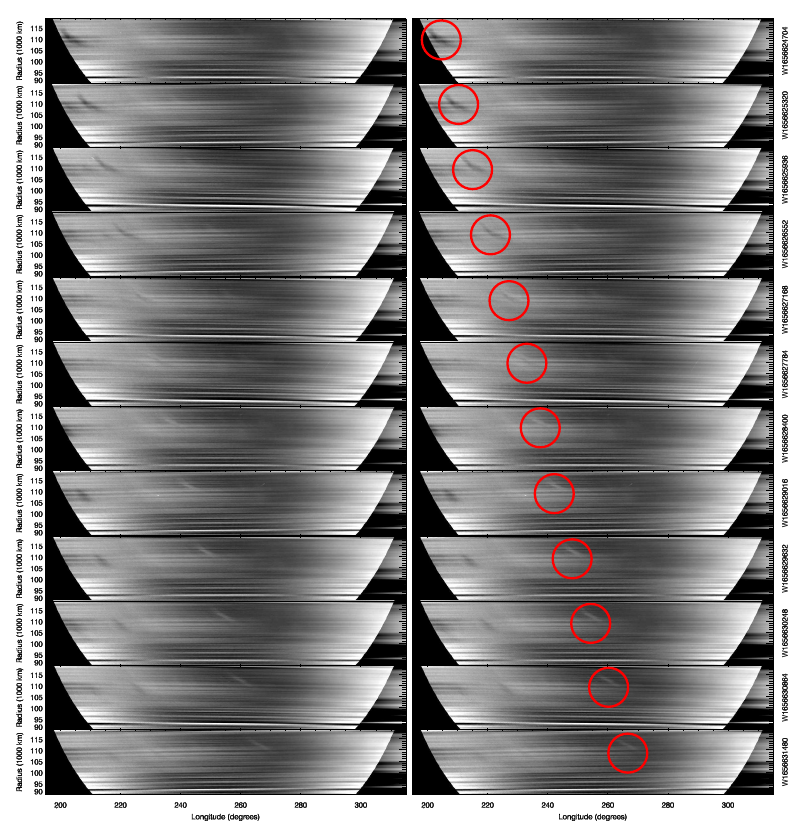}
    \caption{Example of a variable spoke from an observation of the unlit side of the rings (ISS\_134RI\_SPKMVDFHP001\_PRIME), which was obtained {on June 30, 2010} at a phase angle of 110$^\circ$ and a solar elevation angle of 4.9$^\circ$. The two columns show a series of maps (filtered so that the average brightness at each radius has been removed) where a spoke can be observed moving from left to right (marked with circles on the right set of maps). The spoke clearly goes from being darker than the background ring at the start of the sequence to being brighter than the background ring near the end of the sequence.}
    \label{fig: 134_001 transition}
\end{figure*}

\begin{figure*}[tbp]
    \centering
    \includegraphics[width=7in]{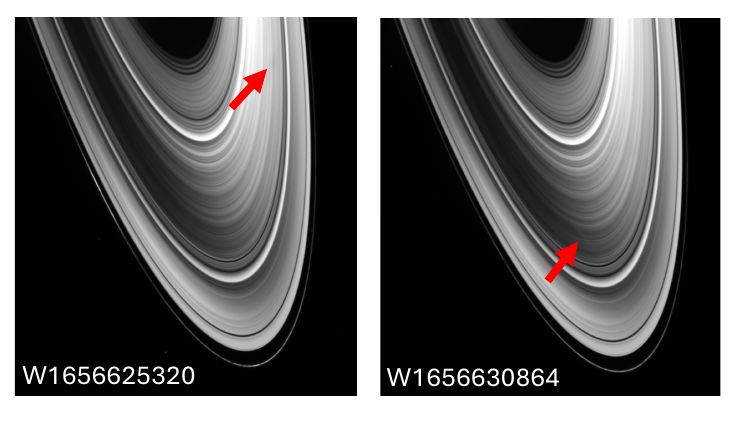}
    \caption{Representative images from the observation of the unlit side of the rings (ISS\_134RI\_SPKMVDFHP001\_PRIME) that produced the maps shown in Figure~\ref{fig: 134_001 transition}. Both images have arrows indicating the location of the spoke marked by the circle in Figure~\ref{fig: 134_001 transition}. Note that the spoke is less obvious in these images than in the maps because no filtering has been used to isolate the spoke signals. Still, the spoke does appear as a subtle dark feature against the relatively bright part of the rings in the left image, and as a bright feature against the darker part of the rings in the right image. }
    \label{fig: 134_001 im}
\end{figure*}

Figure \ref{fig: 134_001 transition} provides a particularly clear example of how the appearance of a single spoke on the unlit side of the rings can vary over time. This figure shows maps derived from images W1656824704 to W1658831480 in observation sequence ISS\_134RI\_SPKMVDFHP001\_PRIME. The spoke enters from the left side of the map and is clearly a dark feature, but as the spoke moves across the ring it transitions to being brighter than its surroundings. {This change is also visible in the original images, as shown in Figure~\ref{fig: 134_001 im}.} These sorts of transitions are relatively common in unlit-side observations with sufficient longitude coverage, and {arise from} the more complicated illumination conditions on the unlit side of the rings, which cause the background ring brightness to vary substantially with longitude.

While the Sun and the background B ring are by far the dominant light sources for spokes on the lit side of the rings, on the unlit side of the rings light scattered from the planet also needs to be taken into account. During the observation shown in Figures  \ref{fig: 134_001 transition} and \ref{fig: 134_001 im}, the longitude of the sun was around 235$^\circ$. This means that around this longitude the rings were exposed to light from the lit side of the planet. Furthermore,  the spacecraft was at a longitude of 338$^\circ$, so as the observed ring longitude increases from 230$^\circ$ to around 300$^\circ$,  the light from the planet scattered by the ring particles goes from being backscattered to being forward scattered. The B-ring is strongly back-scattering, so the background ring appears brighter on the left side of the maps {(which corresponds to the right part of the images shown in Figure~\ref{fig: 134_001 im})}. The dark appearance of the spoke at these longitudes can therefore be understood if we recognize that the planet is the main source of illumination for the background B ring. The spoke is being observed at the equivalent to low phase for that light source, and so the signal is dominated by the spoke scattering light from the background ring away from the camera. However, as the spoke moves around the ring, the background ring brightness declines, and the spoke becomes a bright feature. In this region, the spoke particles could potentially be efficiently forward-scattering light from both the planet and the sun, but in practice it appears that most of the spoke signal comes from sunlight filtering through the rings. Note that when the spoke appeared as a dark feature around 220$^\circ$ it could be easily seen between 105,000 km and 115,000 km. By contrast, when the spoke appears as a bright feature it can only clearly be seen between 110,000 km and 115,000 km. The lack of a visible bright spoke interior to 110,000 km is most easily explained by the fact that the optical depth of the ring undergoes an abrupt transition at 110,000 km, with the normal optical depth being about 3 exterior to this radius and over 5 interior to this radius \citep{Colwell09, HN16}. The amount of sunlight passing through the ring therefore drops by roughly an order of magnitude interior to 110,000 km, and so the inner part of this spoke is likely no longer visible simply because an inadequate amount of sunlight is getting through the rings to be scattered by the spoke particles. 

These general trends can be observed in multiple observation sequences, so the appearance of spokes on the unlit side of the rings does not simply depend on the properties of the spoke particles, but also on the exact radius and longitude where the spokes are actually observed. Since individual spokes can transition from being bright to dark as they move around the planet,  we prefer to avoid categorizing these spokes as either bright or dark, and instead refer to all spokes seen on the unlit side of the rings as ``variable".  While photometrically modeling these spokes will be quite complicated, the observed changes in their contrast should provide novel constraints on spoke particle properties.

\subsection{Mixed spokes}

\begin{figure*}[tbp]
    \centering
    \includegraphics[width=7in]{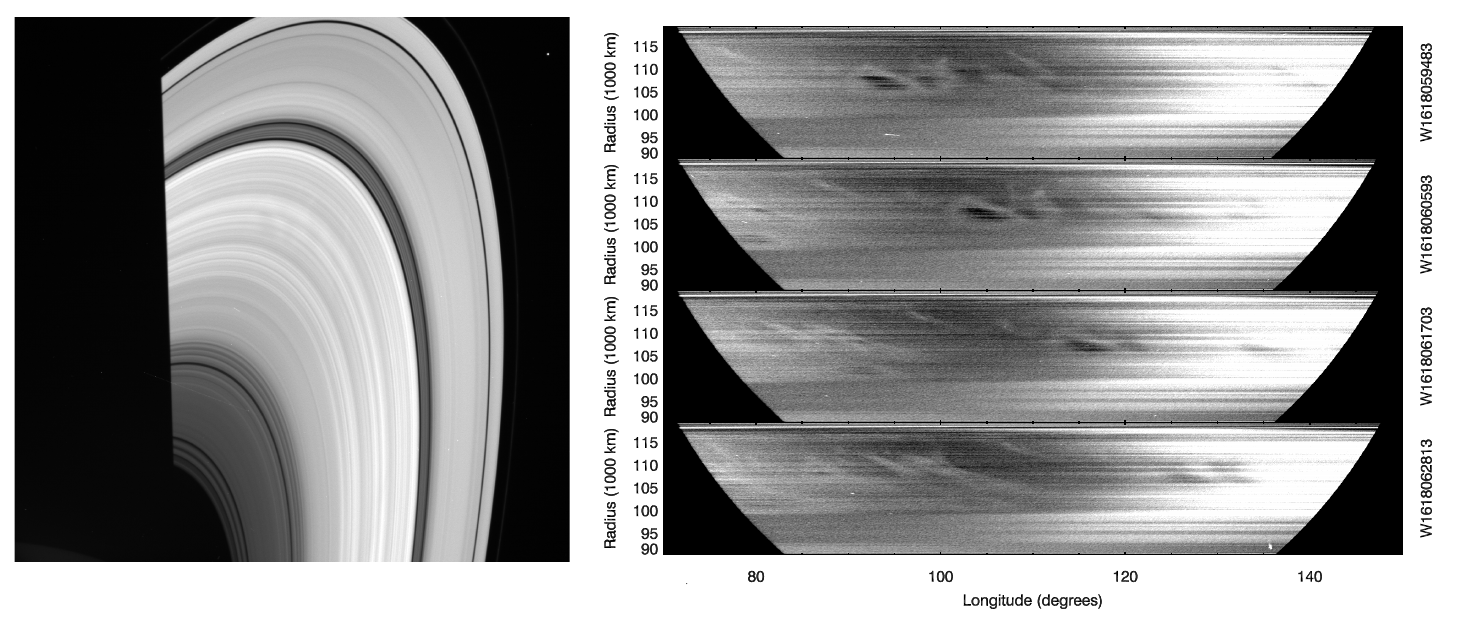}
    \caption{{Mixed spokes seen on the lit side of the rings. The left image shows a representative image (W1618059483)} from the observation sequence ISS\_108RI\_SPKMVLFLP001\_PRIME, which was obtained {on April 10, 2009} at a phase angle of  40$^\circ$ and a solar elevation angle of -2.9$^\circ$. {Faint spokes are visible near the ansa of this image, but they are much less obvious than the spokes in Figures~\ref{fig:074} and~\ref{fig:117_005}.} The right image shows a series of maps derived from this sequence where the average brightness at each radius has been subtracted off to better show the signals from the spokes. These maps reveal that several spokes have dark interiors surrounded by bright edges. }
    \label{fig: 108 mixed}
\end{figure*}

This survey also revealed an unusual type of spoke that to our knowledge has not previously been identified. We will call these features ``mixed spokes" because they have both bright and dark components when observed at roughly the same location. This kind of spoke is most dramatically visible in the observing sequence ISS\_108RI\_SPKMVLFLP001\_PRIME, which is a lit-side spoke observation obtained {on June 14, 2008} at a phase angle of  40$^\circ$ and a solar elevation angle of -2.9$^\circ$. As shown in Figure \ref{fig: 108 mixed}, this sequence contains a spoke whose center is clearly darker than the background  ring, but whose edges are brighter than the background ring.  This basic structure was robust against {various} background subtraction methods.

\begin{figure*}[tbp]
    \centering
    \includegraphics[width=7in]{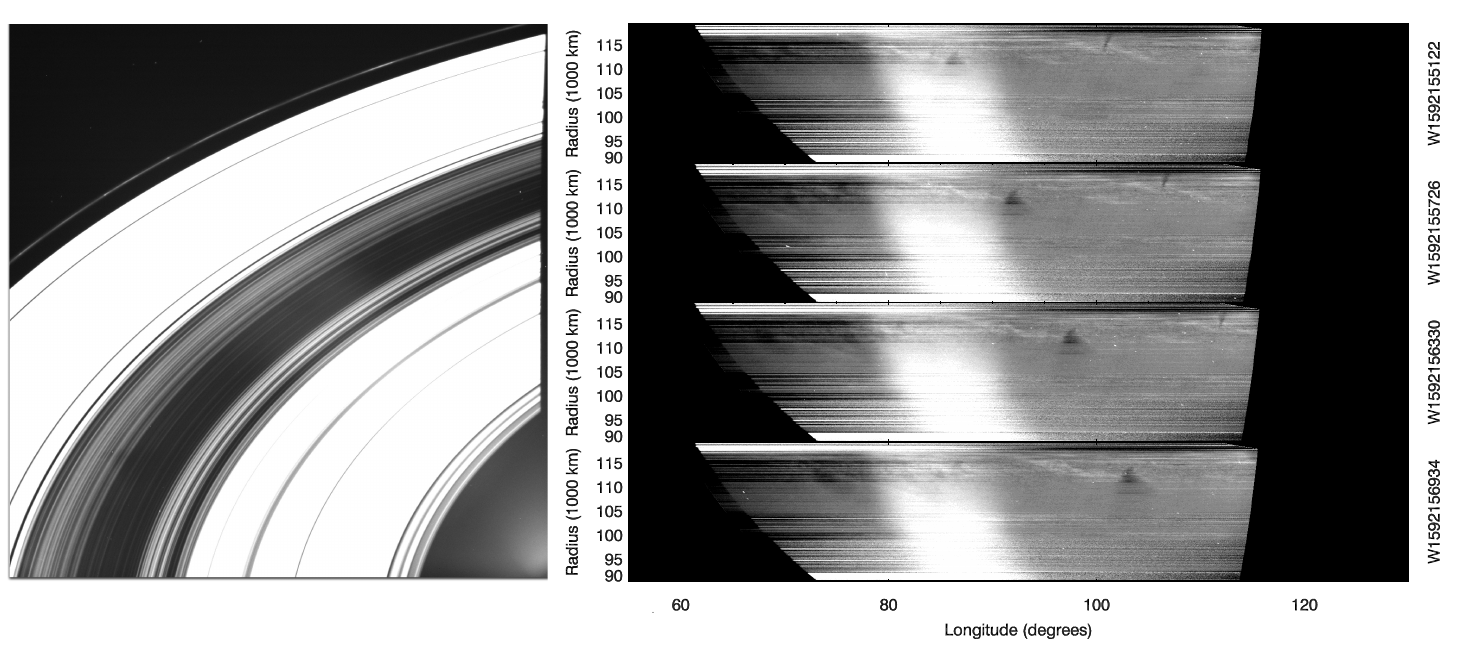}
    \caption{{Mixed spokes seen on the unlit side of the rings. The left image shows a representative image (W1592156934) } from the observation sequence ISS\_072RI\_SPKHRLPDF001\_PRIME,  which was obtained {on June 14, 2008} at a phase angle of  32$^\circ$ and a solar elevation angle of -6.5$^\circ$. {A faint dark spoke is visible at the lower left of this image, but the signals from the spokes are again relatively subtle.} The right image shows a series of maps derived from images in this sequence where the average brightness at each radius has been subtracted off to better show the signals from the spokes. These maps show a dark triangular patch associated with a thin bright structure extending over 10$^\circ$ in longitude.}
    \label{fig: 072 mixed}
\end{figure*}

Unlike the variable spokes discussed above, this unusual spoke appearance cannot be attributed to variations in the background lighting conditions because the basic morphology remains the same as different parts of the spoke rotate around the planet. Instead, this unusual behavior probably represents variations in the particle properties across the spoke itself. Note that these images were obtained when the spacecraft was observing the lit side of the rings at a geometry that falls near the boundary between regions where spokes would appear as bright or dark (see Figure~\ref{timeline}). Thus this image sequence probably corresponds to conditions where the reduction of light from the background B ring nearly balances the enhanced scattering from the spoke itself. Since the ratio of the spoke's opacity to its scattering efficiency depends on its particle properties, particularly their size distribution, the patterns in Figure~\ref{fig: 108 mixed} {are probably due to} variations in the particle size distribution across the spoke. In particular,  smaller particles tend to scatter light over a broader range of angles than larger ones, so the center of the spoke appearing dark while its outskirts appear bright could imply that the typical size of spoke particles decreases with distance from the spoke's center. However, since scattering efficiencies can be complex functions of particle size, shape and composition at these phase angles, more detailed modeling of these spokes is needed to confirm this {hypothesis}. Such studies will provide important information about trends in particle properties that could have implications for spoke formation and evolution.

ISS\_108RI\_SPKMVLFLP001\_PRIME is the only observation sequence to show clear mixed spokes on the lit side of the rings. However, mixed spokes can also be found in a number of observations of the unlit side of the rings.  A particularly clear example is shown in Figure~\ref{fig: 072 mixed}. In this case, we see a dark triangular patch at the end of a long, thin spoke extending over 10$^\circ$ in longitude. These two features move through the same longitudes so the difference in contrast between the two features cannot be due to changing illumination conditions, but instead must again reflect some difference in the light-scattering properties of these two structures. In this case, we can probably regard these two features as two distinct spokes {that} have different particle properties. Nevertheless they are still interesting as an opportunity for modeling to quantify particle size variations among different spokes.

\section{Quantifying variations in spoke intensity}
\label{quantify}

The standardized B-ring maps shown in the previous section not only provide qualitative insights into spoke properties, but can also yield quantitative information about trends in spoke activity. While detailed analyses of individual spoke properties are beyond the scope of this particular paper, we can use relatively simple statistics derived from these maps to chart trends in overall spoke activity over the course of the Cassini mission. 

In Section~\ref{extremes} below we identify the earliest and latest observations of spokes obtained by Cassini that can be identified in our maps, and use these to illustrate the limitations of considering the presence or absence of spokes as a simple binary choice. We therefore develop a more quantitative metric of spoke activity in Section~\ref{metrics} and compute this metric for a sub-set of the lit-side observations in Section~\ref{trends} in order to document trends in spoke activity over time.

\pagebreak

\subsection{Search for Earliest and Last Cassini Spoke Detections}
\label{extremes}

\begin{figure*}[tbp]
    \centering
    \includegraphics[width=6.5in]{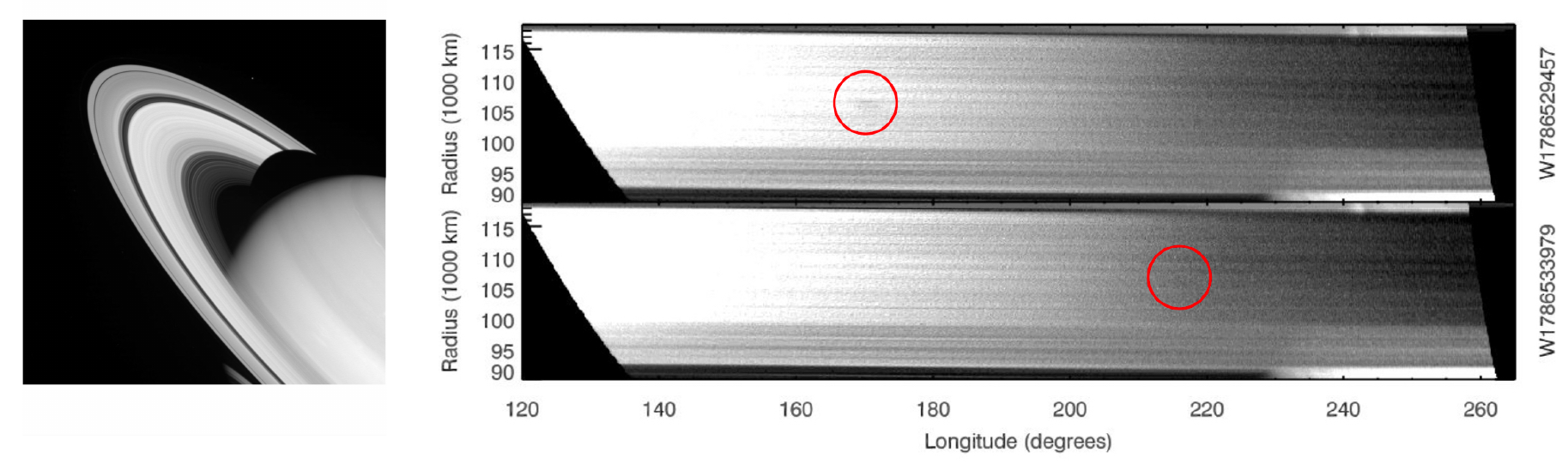}
    \caption{The last spoke  seen on the lit side of the rings from the Cassini mission, from observation sequence IOSIC\_207RC\_COMPLLRES001\_SI, which was obtained {on August 12, 2014} at a phase angle of 16.7$^\circ$ and a solar elevation angle of 22.7$^\circ$. {This faint dark spoke cannot be easily seen in the sample image (W1786529457) shown on the left, but } can be seen in the two filtered maps at the right at the locations marked by circles.}
    \label{fig: final spoke1}
\end{figure*}

\begin{figure}[btp]
    \centering
    \includegraphics[width=3.25in]{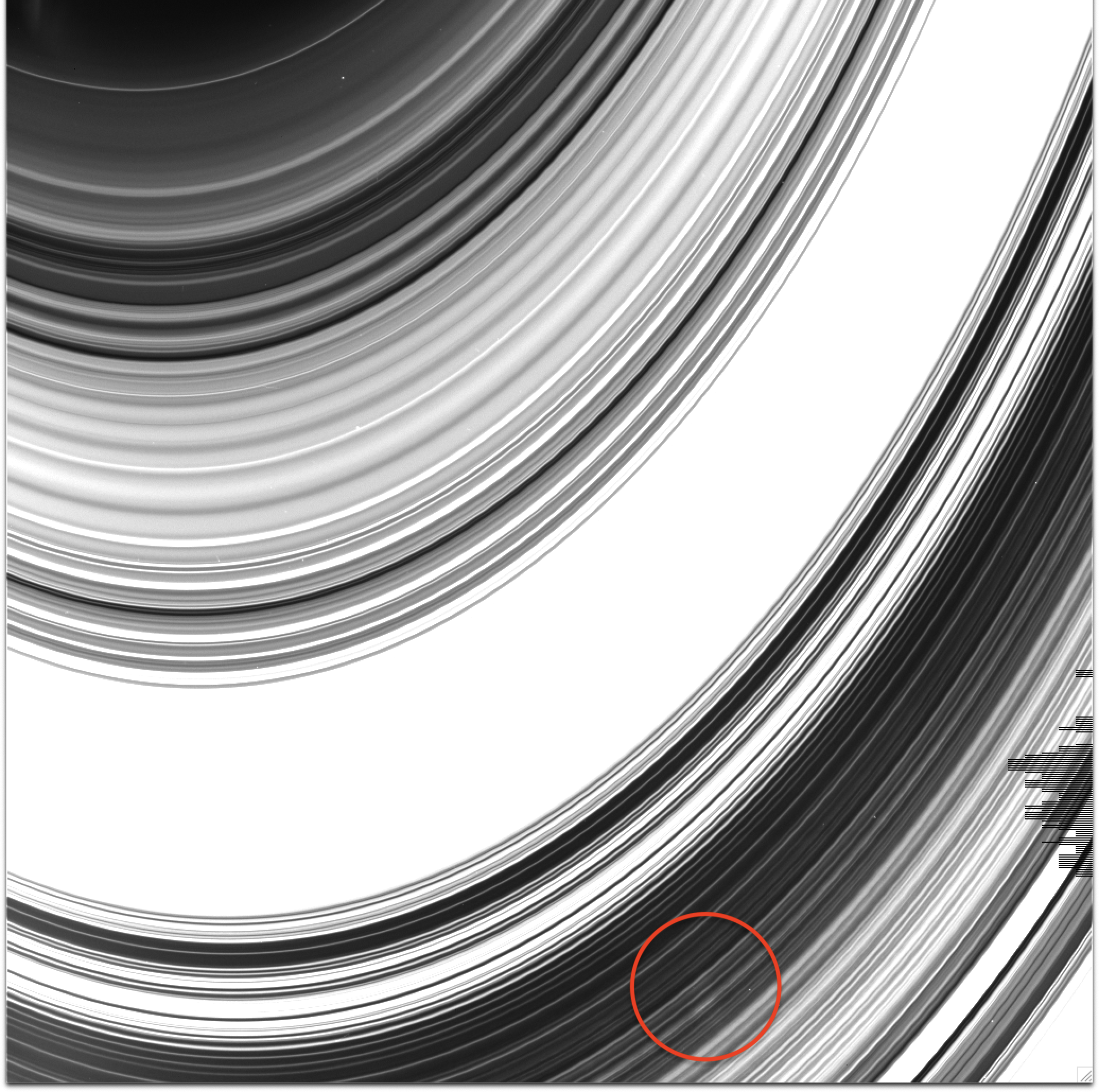}
    \caption{The last spoke seen on the unlit side of the rings from the Cassini mission, from an image {(W1870770784)} in observation sequence ISS\_269RI\_HIPHASEFB001\_PIE, which was obtained {on April 13, 2017} at a phase angle of 175$^\circ$ and a solar elevation angle of 26.7$^\circ$. The spoke is the bright feature marked by the circle in this particular image.}
    \label{fig: final spoke2}
\end{figure}

While we were performing the initial examination of the ring observations, we also conducted a visual search for faint spokes near the beginning and end of the Cassini mission. The earliest reported spokes are those from observation  IOSIC\_014RI\_SUBMU12HP001\_SI \citep{Mitchell06}, which were obtained {on September 5, 2005} at a solar elevation angle of -20.4$^\circ$ and observed the unlit side of the rings. Thus far, we have not found any convincing spoke signals prior to this particular observation, although there may be some potential spoke-like patches near the noise limit in certain images.

On the other hand, the latest previously identified spokes were  from the high-phase, unlit-side observation sequence  IOSIC\_206RI\_SP160L7001\_SI, which were obtained {on July 17, 2014} at a solar elevation angle of  22.5$^\circ$ and showed bright spokes on the unlit side of the rings \citep{Simon23}. Our search through the available Cassini images revealed two later observations that contained faint but identifiable spokes. 

First, the  low-phase, lit-side observation IOSIC\_207RC\_COMPLLRES001\_SI (obtained {on August 12, 2014} at a solar elevation angle of 22.7$^\circ$) contained a single faint dark spoke. This spoke was not obvious in the images, but is clearly present in two filtered maps derived from images W178653979 and W178529457 (see Figure \ref{fig: final spoke1}). While this feature is faint in both images, it is most likely a real spoke because in the 1.25 hours between those two images the feature moved roughly 45$^\circ$  in longitude, which is consistent with the expected orbital mean motion of B-ring material.

The other late spoke we identified was part of the observation ISS\_269RI\_HIPHASEFB001\_PIE, which was a mosaic of the unlit side of the ring system obtained near the end of the Cassini mission {on April 13, 2017} at a solar elevation angle of 26.7$^\circ$ at exceptionally high phase angles of 175$^\circ$.  A set of five images that contained the B ring showed a faint bright spoke (see Figure~\ref{fig: final spoke2}). 

These late spokes demonstrate an issue with using the first and last appearance of spokes to document the seasonality of spokes. While this approach is sensible for earth-based observations that have a limited range of viewing/lighting geometries, defining spoke appearance and disappearance becomes more challenging when dealing with the broad range of viewing geometries observed by Cassini. More fundamentally, our survey showed that the seasonality of spokes could not simply be expressed by spokes being present or absent. Close to equinox, spokes can easily be seen in unprocessed images, while early and late in the Cassini mission, spokes can still be detected but are generally fainter and take more effort to identify. For this reason, we decided to develop a way to quantify the average overall spoke activity in a given observation.

\subsection{Spoke activity metrics for lit-side images}
\label{metrics}

Most of the prior efforts to quantify spoke activity used qualitative categories \citep{Porco82, Porco83, Grun83, Mitchell13}, and more complex techniques are currently being developed to automatically identify, count and quantify individual spokes within each image \citep{Byrne23}.  For this study we are more interested in the long-term variations in overall spoke activity,  so we will build upon prior work by \cite{McGhee05} and consider statistics that can provide comparable estimates of the average spoke intensity for a reasonably broad range of lighting and viewing conditions. As a practical matter, we will here focus exclusively on spokes observed on the sunlit side of the rings, since the variable appearance of spokes on the unlit side of the rings complicates any effort to quantify the typical spoke intensity.

\cite{McGhee05} quantified the brightness of individual spokes in terms of the difference in reflectance ($I/F$) between the spoke and the background B ring. Generalizing this concept, we define a quantity we call the ``spoke signal'' $s(r)$ as the $rms$ brightness variations of the rings at each sampled radius $r$ in the re-projected maps:

\begin{equation}
s(r) = \sqrt{\left\langle\left(I/F_s(r, \lambda) - \overline{I/F}_s(r)\right)^2\right\rangle}
\label{eq1}
\end{equation}
where $I/F_s$ are the observed $I/F$ within the map (which can vary both with radius $r$ and  with longitude $\lambda$), while $\overline{I/F}_s$ is the average brightness of the ring at the relevant radius. As we will see below, for the lit-side observations considered here, $\overline{I/F}_s$ is a good approximation of the background ring brightness in the absence of spokes, which we will designate as $I/F_b$.  These $rms$ variations are straightforward to compute and should provide a reasonable measure of spoke intensity because when either bright or dark spokes are present in the image, they should increase the dispersion of the ring's brightness over the range of radii containing the spokes. 

To better understand the relationship between $s(r)$ and physical properties of the spokes, consider a location in the rings where a spoke with a normal optical depth $\tau_n$ is sitting above a part of the ring with background brightness $I/F_b$. If the spoke is also being viewed at an emission angle $e$, then the optical depth of the spoke along the line of sight  $\tau=\tau_n/\cos(e)$. So long as the spoke is on the sunlit side of the rings (so that illumination from the planet can be neglected) and the spoke optical depth is low enough \citep[so that multiple scattering can be neglected, cf.][]{McGhee05}, the spoke will have two different effects on the observed ring brightness. First, it will scatter some of the light from the background ring away from the line of sight, which will reduce the ring's brightness from $I/F_b$ to $TI/F_b$, where $T=e^{-\tau}$ is the transmission through the spoke material. Second, the spoke itself will scatter some additional light from the Sun into the camera.  If we assume the spoke is sufficiently optically thin that the transmission through the spoke $T\simeq (1-\tau)$ and the brightness of the spoke material is directly proportional to $\tau$, then the observed $I/F$ of the ring is given by the following expression:

\begin{equation}
I/F_s=I/F_b\left[1-\frac{\tau_n}{\cos(e)}\right]+ B(\alpha)\frac{\tau_n}{\cos(e)}
\label{eq2}
\end{equation}
where $B(\alpha)$ is the appropriately normalized scattering efficiency of the spoke material at the observed phase angle $\alpha$. In general, $B(\alpha)$ depends on the particle size distribution in the spokes and the existence of mixed spokes indicates that this parameter can vary with location. However, given that mixed spokes are very rarely observed on the lit side of the rings, in practice we expect that the observed brightness variations on the ring are predominantly due to variations in $\tau_n$. Hence, for the sake of simplicity we will here assume $B(\alpha)$ {has approximately the same value everywhere in all} the spokes seen in each observation. 

So long as $B$ is approximately constant, then inserting Equation~\ref{eq2} into Equation~\ref{eq1} reveals that $s$ is directly proportional to the rms variations in the spoke normal optical depth $\tau_n$.

\begin{equation}
s(r)=\frac{I/F_b}{|\cos(e)|}\left|1-\frac{B(\alpha)}{I/F_b}\right|\sqrt{\langle(\tau_n-\bar{\tau}_n)^2\rangle}
\end{equation}
where $\bar{\tau}_n$ is the average spoke optical depth at each radius. Note that both $I/F_b$ and $\cos(e)$ do not depend on the properties of the spokes themselves and $I/F_b$ is well approximated as the average ring brightness at each radius $\overline{I/F}_s$ (see below), while $\cos(e)$ is derived from the observation geometry. Hence we can define a ``Normalized Spoke Signal'' $n(r)$ as:

\begin{equation}
n(r)=\frac{|\cos(e)|}{I/F_b}s(r) =\left|1-\frac{B(\alpha)}{I/F_b}\right|\sqrt{\langle(\tau_n-\bar{\tau_n})^2\rangle}
\label{neq}
\end{equation}
 In the limit where $B(\alpha)<<I/F_b$,  $n(r)$ is equivalent to the $rms$ variations in the spoke optical depth, which should be independent of viewing or lighting geometry.\footnote{Note also that this quantity differs from the generalization of the normalized spoke contrast used by \cite{McGhee05} by a factor of $|\cos(e)|$.} This quantity is therefore a sensible statistic for quantifying the prevalence of dark spokes on the lit side of the rings. This will not be the case when $B(\alpha)\gtrsim I/F_b$, which corresponds to situations when the spokes are bright and/or the background ring is dark. However, even in those situations, the constant of proportionality between $n(r)$ and the $rms$ optical depth variations should depend only on the observed phase angle, so this quantity still facilitates comparisons among the various  observations. Furthermore, in practice we find that $n(r)<<1$ for all the observations considered in this initial study,  justifying our assumption that  $\overline{I/F}_s \simeq I/F_b$.

\begin{figure}[tp]
    \centering
    \includegraphics[width=5in]{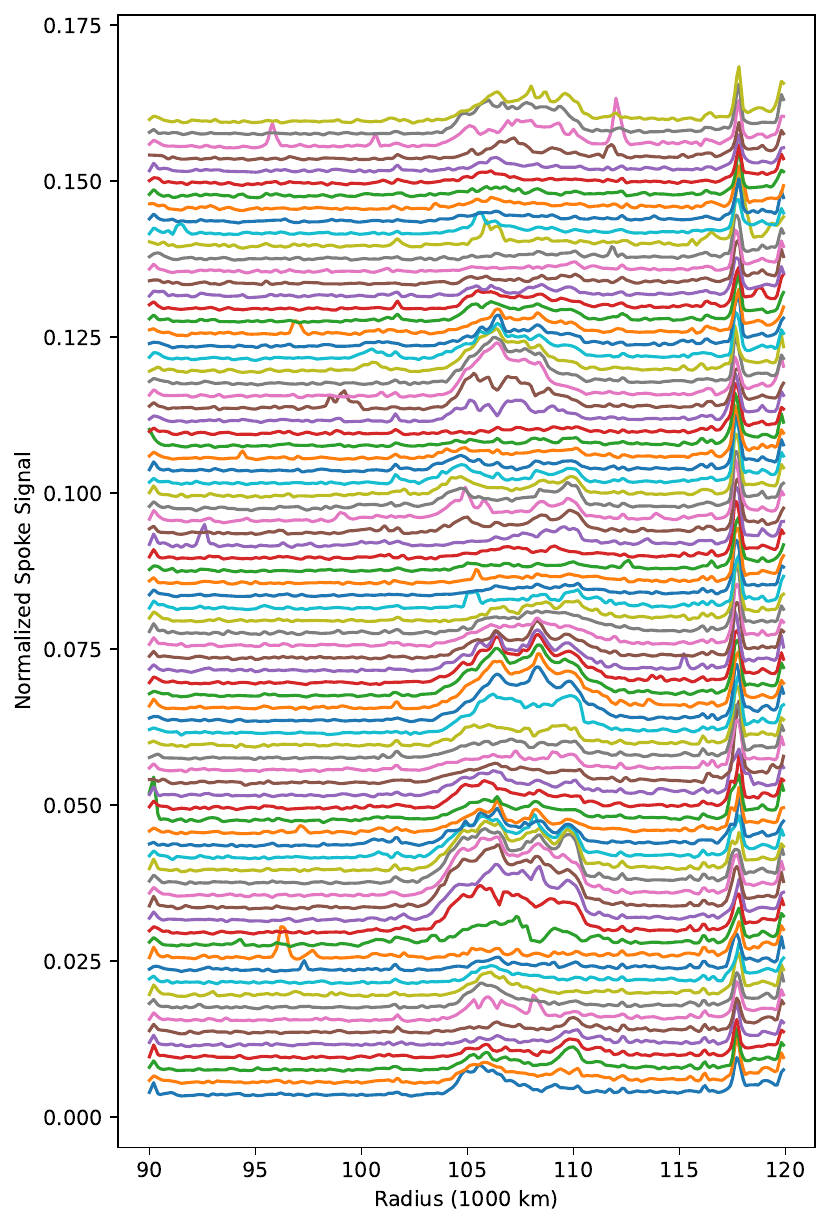}
    \caption{Normalized spoke signal profiles derived from the individual maps from the ISS\_117RI\_SPKMVLFHP005\_PRIME (see Figure~\ref{fig:117_005}). Each curve gives the value of $n$ as a function of radius for a single image. The different profiles have been vertically offset for clarity. The peak around 117,000 km corresponds to the edge of the B ring, and the variable bumps between 105,000 km and 110,000 km are due to the spokes in this region.}
    \label{fig:117 indv signals}
\end{figure}

In practice, we compute $s(r)$ and $I/F_b$ from each of the individual maps using a python code that takes the brightness data at each radius, selects out a longitude range, and computes the mean brightness to obtain an estimate of $I/F_b$.  In order to estimate $s(r)$, these same data are fit to a quadratic function of longitude and the $rms$ variations of the residuals to that fit are computed. {Removing the quadratic trend reduces the contamination of $s(r)$ due to changes in the lighting geometry or slight pointing errors}.  

 Figure \ref{fig:117 indv signals} shows examples of the resulting profiles from the same observation with strong bright spokes shown in Figure~\ref{fig:117_005}. Note that all the profiles show a peak around 117,000 km that can be attributed to the sharp outer edge of the B ring. More importantly, these profiles show variable bumps between 103,000 and 112,000 km.  The locations of these bumps are consistent with the observed location of the spokes (see Figure \ref{fig:117_005}), {demonstrating} that this statistic provides useful information about spoke activity. 

\begin{table}[!ht]
\caption{Lit-Side observations used to quantify spoke activity}
\label{tab:profobs}
 \hspace{-1in}   \resizebox{7.5in}{!}{\begin{tabular}{|l|l|l|l|l|l|l|l|l|l|l|}
    \hline
        Observation Sequence & Type$^a$ & Date$^b$ & Solar El. & Emission & Phase & No. & Longitude & ${I/F_b} ^g$ & $N^h $ & ${N’}^i$ \\ 
        &  & & Angle$^c$ ($^\circ$) & Angle$^d$ ($^\circ$) & Angle$^e$ ($^\circ$) & Images & Range$^f$ ($^\circ$) & & $(10^{-3})$ & $(10^{-3})$ \\ \hline
        ISS\_00BRI\_SPKMOVPER002\_PRIME & X & 2004-326 & -23.34 & 101.9 & 76.1 & 87 & 180-190 & 0.19 & 0.14 & 0.17 \\ \hline
        ISS\_007RI\_SPKLRSLPA002\_PRIME & X & 2005-116 & -21.88 & 105.45 & 50.7 & 4 & 40-80 & 0.225 & 0.11 & 0.1 \\ \hline
        ISS\_063RI\_SPKFORMLF001\_PRIME & D & 2008-094 & -7.63 & 104.41 & 17.1 & 7 & 60-105 & 0.313 & 0.74 & 0.73 \\ \hline
        ISS\_066RI\_SPKSTFORM001\_PRIME & D & 2008-123 & -7.18 & 93.68 & 11.3 & 4 & 90-130 & 0.518 & 0.51 & 0.44 \\ \hline
        ISS\_067RI\_SPKFORMLF001\_PRIME & D & 2008-132 & -7.04 & 109.73 & 22.7 & 7 & 60-100 & 0.212 & 0.46 & 0.65 \\ \hline
        ISS\_071RI\_SPOKEMOV001\_PRIME & D & 2008-162 & -6.58 & 108.35 & 19.1 & 17 & 80-140 & 0.215 & 0.4 & 0.34 \\ \hline
        ISS\_074RI\_SPKLFMOV001\_PRIME & D & 2008-184 & -6.25 & 97.99 & 14.4 & 17 & 80-140 & 0.374 & 0.67 & 0.56 \\ \hline
        ISS\_078RI\_SPKFORM001\_PRIME & D & 2008-211 & -5.83 & 127.76 & 36.8 & 69 & 70-110 & 0.106 & 3.56 & 3.71 \\ \hline
        ISS\_081RI\_SPKMVLFLP001\_PRIME & D & 2008-234 & -5.48 & 109.19 & 23.1 & 40 & 80-130 & 0.173 & 2.01 & 1.59 \\ \hline
        ISS\_085RI\_SPKMVLFLP001\_PRIME & D & 2008-263 & -5.01 & 104.39 & 21.5 & 34 & 90-140 & 0.209 & 1.45 & 1.31 \\ \hline
        ISS\_086RI\_SPKMVLFLP002\_PRIME & D & 2008-271 & -4.91 & 108.89 & 24.6 & 35 & 90-140 & 0.156 & 1.94 & 1.48 \\ \hline
        ISS\_088RI\_SPKMVLFLP001\_PRIME & D & 2008-286 & -4.68 & 103.6 & 21.9 & 46 & 90-140 & 0.213 & 1.28 & 0.97 \\ \hline
        ISS\_094RI\_SPKMVLFLP001\_PRIME & D & 2008-331 & -3.98 & 122.86 & 38.8 & 49 & 80-120 & 0.068 & 5.53 & 5.42 \\ \hline
        ISS\_096RI\_SPKFMLFLP001\_PRIME & D & 2008-246 & -3.74 & 121.18 & 68.6 & 5 & 60-110 & 0.012 & 4.89 & 4.77 \\ \hline
        ISS\_102RI\_SPKFMLFLP001\_PRIME & D & 2009-036 & -2.87 & 118.52 & 40.5 & 9 & 90-110 & 0.048 & 4.04 & 4.00 \\ \hline
        ISS\_108RI\_SPKMVLFLP001\_PRIME & M & 2009-100 & -1.89 & 111.55 & 39.9 & 19 & 90-130 & 0.042 & 1.01 & 0.78 \\ \hline
        ISS\_109RI\_SPKFMLFHP001\_PRIME & B & 2009-112 & -1.7 & 144.46 & 123.5 & 5 & 80-90 & 0.006 & 2.38 & 2.02 \\ \hline
        ISS\_117RI\_SPKMVLFHP001\_PRIME & B & 2009-242 & 0.31 & 78.35 & 78.4 & 72 & 180-240 & 0.013 & 1.84 & 2.58 \\ \hline
        ISS\_117RI\_SPKMVLFHP003\_PRIME & B & 2009-243 & 0.32 & 78.86 & 84.5 & 110 & 180-260 & 0.013 & 2.99 & 2.3 \\ \hline
        ISS\_117RI\_SPKMVLFHP004\_PRIME & B & 2009-245 & 0.34 & 79.33 & 89.2 & 85 & 180-260 & 0.013 & 2.63 & 2.11 \\ \hline
        ISS\_117RI\_SPKMVLFHP005\_PRIME & B & 2009-246 & 0.35 & 79.72 & 92.5 & 79 & 180-270 & 0.012 & 3.15 & 2.71 \\ \hline
        IOSIC\_124RI\_EQLBN002\_SI & B & 2010-011 & 2.35 & 80.7 & 56.9 & 5 & 140-190 & 0.059 & 4.75 & 3.17 \\ \hline
        IOSIC\_132RI\_EQLBN001\_SI & B & 2010-151 & 4.44 & 80.61 & 115.1 & 4 & 240-280 & 0.038 & 1.52 & 0.85 \\ \hline
        IOSIC\_132RI\_EQLBN003\_SI & B & 2010-153 & 4.47 & 79.32 & 133.7 & 5 & 100-120 & 0.032 & 1.1 & 0.47 \\ \hline
        IOSIC\_132RI\_EQLBN004\_SI & B & 2010-153 & 4.47 & 79.32 & 133.7 & 4 & 260-300 & 0.03 & 2.48 & 1.72 \\ \hline
        IOSIC\_134RI\_P50L30S15001\_SI & D & 2010-186 & 4.96 & 72.77 & 19.6 & 4 & 120-180 & 0.174 & 1.67 & 1.57 \\ \hline
        IOSIC\_134RI\_P50L30S15002\_SI & D & 2010-186 & 4.97 & 80.1 & 19 & 7 & -20-20 & 0.244 & 0.65 & 0.49 \\ \hline
        ISS\_168RB\_BMOVIE001\_PRIME & D & 2012-182 & 14.88 & 82.2 & 15.9 & 6 & 160-200 & 0.486 & 0.19 & 0.14 \\ \hline
        ISS\_199RI\_SPOKEMOV002\_PRIME & B & 2013-315  & 20.34 & 77.32 & 144 & 122 & 180-220 & 0.073 & 0.18 & 0.16 \\ \hline
        ISS\_199RI\_SPOKEMOV003\_PRIME & B & 2013-317 & 20.35 & 73.18 & 138.4 & 150 & 180-220 & 0.073 & 0.29 & 0.18 \\ \hline
        ISS\_199RI\_SPOKEMOV004\_PRIME & B & 2013-319  & 20.37 & 67.35 & 131.1 & 118 & 180-220 & 0.074 & 0.49 & 0.14 \\ \hline
        ISS\_199RI\_SPOKEMOV006\_PRIME & B & 2013-322 & 20.4 & 59.28 & 120.1 & 163 & 180-220 & 0.079 & 0.65 & 0.16 \\ \hline
        ISS\_200RI\_SPOKEMOV001\_PRIME & B & 2013-352 & 20.68 & 68.34 & 133.4 & 55 & 180-220 & 0.073 & 0.52 & 0.21 \\ \hline
        ISS\_200RI\_SPOKEMOV004\_PRIME & B & 2013-357 & 20.73 & 47.65 & 104.5 & 108 & 170-210 & 0.088 & 0.91 & 0.21 \\ \hline
        ISS\_200RI\_SPOKEMOV011\_PRIME & B & 2014-014 & 20.93 & 75.17 & 141.8 & 64 & 180-220 & 0.073 & 0.35 & 0.17 \\ \hline
        ISS\_201RI\_SPOKEMOV011\_PRIME & X & 2014-045 & 21.21 & 61.77 & 124 & 66 & 160-200 & 0.076 & 0.65 & 0.13 \\ \hline
        ISS\_201RI\_SPOKEMOV013\_PRIME & X & 2014-047 & 21.23 & 55.2 & 114.4 & 60 & 160-200 & 0.08 & 0.43 & 0.25 \\ \hline
        ISS\_201RI\_SPOKEMOV015\_PRIME & X & 2014-049 & 21.55 & 49.58 & 105.1 & 66 & 160-210 & 0.087 & 0.47 & 0.23 \\ \hline
        IOSIC\_207RC\_COMPLLRES001\_SI & D & 2014-224 & 22.7 & 73.17 & 16.7 & 5 & 150-200 & 0.499 & 0.16 & 0.38 \\ \hline
    \end{tabular}}
    
\footnotesize    $^a$ Type of spokes visible in the images. X = No spokes visible, D = Dark spokes, B = Bright Spokes, M = Mixed Spokes, V = Variable Spokes (all spokes on dark side of the rings).
    
    $^b$ Data in format Year followed by day of year.
    
    $^c$ Solar Elevation Angle relative to the ring plane, which is negative if the Sun is on the south side of the rings and positive if the Sun is on the north side of the rings
    
    $^d$ Emission Angle, which is the angle between the line of sight to the spacecraft and the ring surface normal. Emission angles below 90$^\circ$ mean the spacecraft views the north side of the rings and emission angles above 90$^\circ$ mean the spacecraft views the south side of the rings.
    
    $^e$ Phase Angle, which is the angle between the incident sunlight and the emitted reflected light from the rings. 
    
    $^f$ Inertial longitude range used to compute the spoke signal (longitude measured relative to the ascending node of the rings on the J2000 reference plane.
    
    $^g$ Average ring brightness between 102,000 km and 112,000 km.
    
    $^h$ Average value of the normalized spoke signal $\bar{n}(r)$ between  102,000 km and 112,000 km above a background based on the average value of $\bar{n}(r)$ in the regions 99,000-100,000 km and 114,000-115,000 km. 
    
        $^i$ Average value of the excess normalized spoke signal $\bar{n}'(r)$ between  102,000 km and 112,000 km above a background based on the average value of $\bar{n}'(r)$ in the regions 99,000-100,000 km and 114,000-115,000 km. 
    
\end{table}

\subsection{Trends in spoke activity for the lit-side observations}
\label{trends}

\begin{figure}[tp]
    \centering
    \includegraphics[width=6.5in]{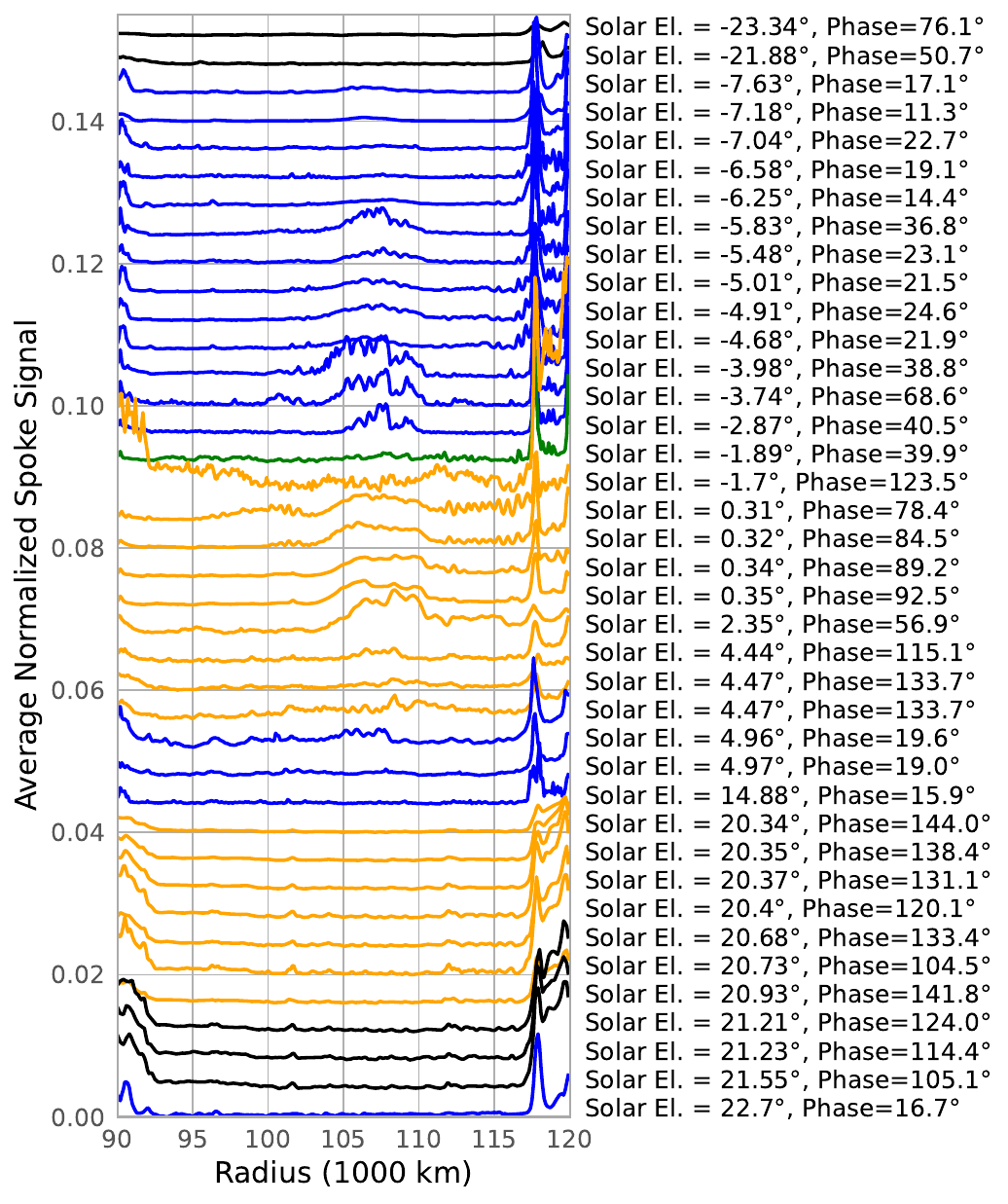}
    \caption{The averaged normalized spoke signal ($\bar{n}$) profiles for lit-side observations containing at least 4 images targeted at the same region of the rings. Curves are sorted by time/solar elevation angle, and have been offset so the minimum values in these profiles are evenly spaced. These curves are also color coded based on the properties of observed spokes. Black curves are observations where no spoke was visible, while blue, orange and green curves are observations where the spokes were dark, bright, or mixed, respectively  Note the clear increase in the signal between 103,000 km and 112,000 km {for solar elevation angles within a few degrees of zero}.}
    \label{fig:profs_avdat}
\end{figure}

In principle, we can assess trends in spoke activity by comparing individual $n(r)$ profiles. However, for this initial study we found it was more practical to instead consider the average value of the $n(r)$ for all {of} the images in each of the lit-side observations that included over 4  wide-angle-camera images of the entire B ring covering the same range of inertial longitudes. Table~\ref{tab:profobs} summarizes the salient properties of this subset of the spoke observations, along with the range of longitudes used to compute the $n(r)$ profiles, while Figure~\ref{fig:profs_avdat} shows the average normalized spoke signal profiles $\bar{n}(r) $ derived from these observations.

Many of the profiles in Figure~\ref{fig:profs_avdat} obtained close to equinox (i.e. solar elevation angles close to zero) show bumps around 105,000-110,000 km indicative of spoke activity. However, some of these profiles (particularly those obtained at higher phase angles) also show structures elsewhere in the rings. Furthermore, these profiles often show substantial constant offsets (these are removed from Figure~\ref{fig:profs_avdat} for the sake of clarity). Both of these aspects of the $\bar{n}(r)$ profiles can be attributed to other sources of brightness variations like stray light and instrumental noise. Since these background trends are largely common to all the images, we found that we could remove most of them by instead computing the difference between the average and the minimum values of $n(r)$ among the profiles for each observation (making the assumption that the minimum $n(r)$ values correspond to no spokes, see below). These profiles are shown in Figure~\ref{fig:profs_avmin} and will be referred to as the average {\sl excess} normalized spoke signal $\bar{n}'(r)$. 

\begin{figure}[tp]
    \centering
    \includegraphics[width=6.5in]{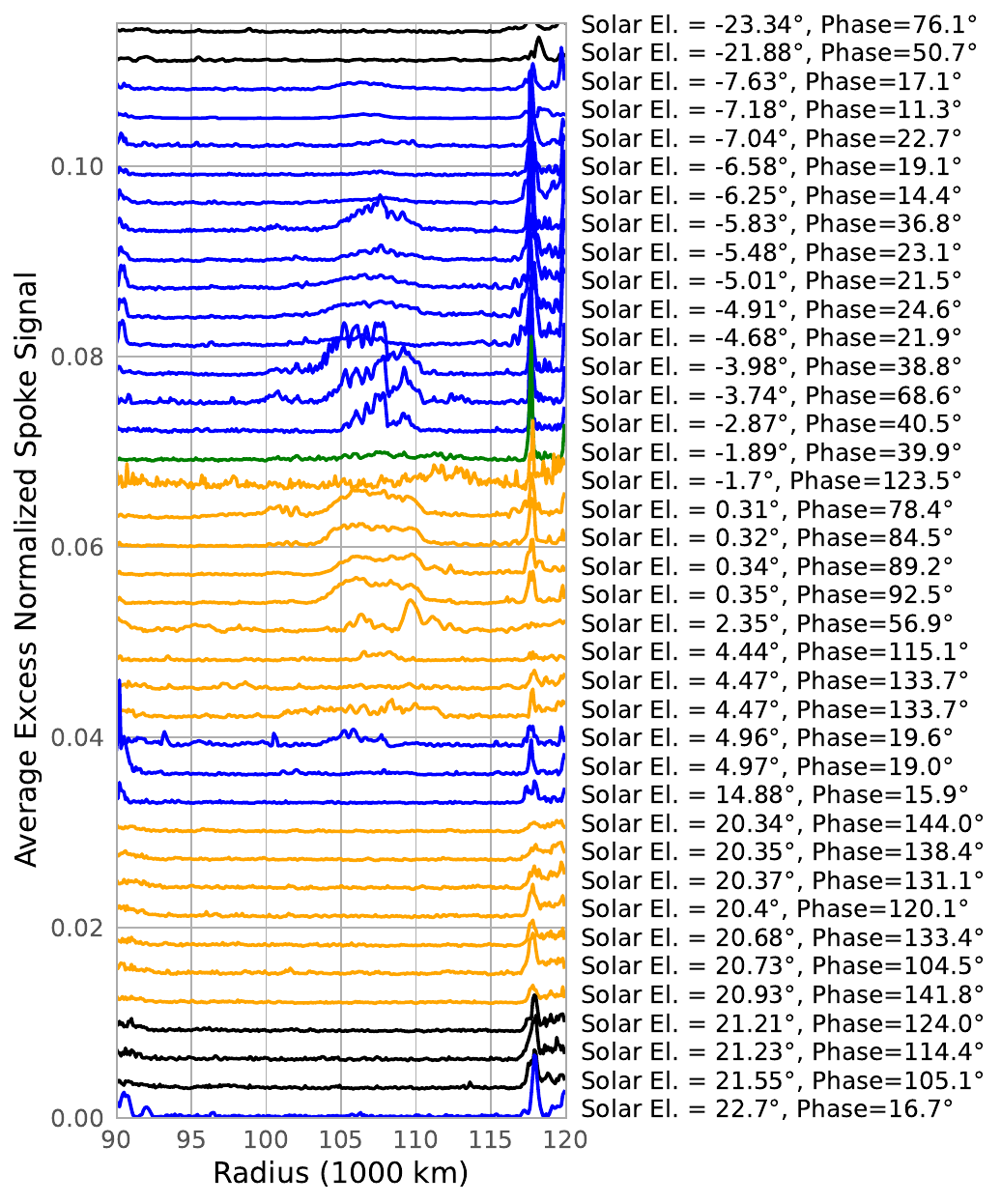}
    \caption{The average excess normalized spoke signal ($\bar{n}'$) profiles for lit-side observations containing at least 4 images targeted at the same region of the rings. Curves are sorted by time/solar elevation angle, and are again color coded based on the properties of observed spokes. Black curves are observations where no spoke was visible, while blue, orange and green curves are observations where the spokes were dark, bright, or mixed.  Compared to Figure~\ref{fig:profs_avdat}, these profiles show less structure outside of the region containing the spokes.}
    \label{fig:profs_avmin}
\end{figure}

The shapes of most of the profiles in Figures~\ref{fig:profs_avdat} and~\ref{fig:profs_avmin} are very similar. In general, the $\bar{n}'(r)$ profiles show less structure  interior to 102,000 km. These profiles also have smaller overall offsets. Both of these results demonstrate that most of those signals are indeed due to common-mode variations among the various images and so are not representative of  spoke-like ring structures. Furthermore, the shapes and magnitudes of the peaks between 100,000 km and 115,000 km are mostly the same in the two sets of profiles, which indicates that subtracting the minimum $n(r)$ does not strongly affect the actual signal from the spokes for most of these observations. This is because even in observations where the spokes are quite active, there are often times when no spokes are present at any given radius, and so the minimum $n(r)$ is close to the appropriate background level (see Figure~\ref{fig:117 indv signals}). The one obvious exception to this behavior is the observation obtained at a solar elevation angle of +2.35$^\circ$, where the $\bar{n}'(r)$ profile has a much lower signal level between 105,000 km and 109,000 km than the corresponding $\bar{n}(r)$ profile. This particular observation only includes 4 images, all of which show spoke activity in that radial range, and thus demonstrates that $\bar{n}'(r)$ can underestimate spoke activity in certain circumstances. While this only happens rarely, we will consider both $\bar{n}(r)$ and $\bar{n}'(r)$ in the following discussion.

\begin{figure*}[p]
    \centering
    \includegraphics[width=6.5in]{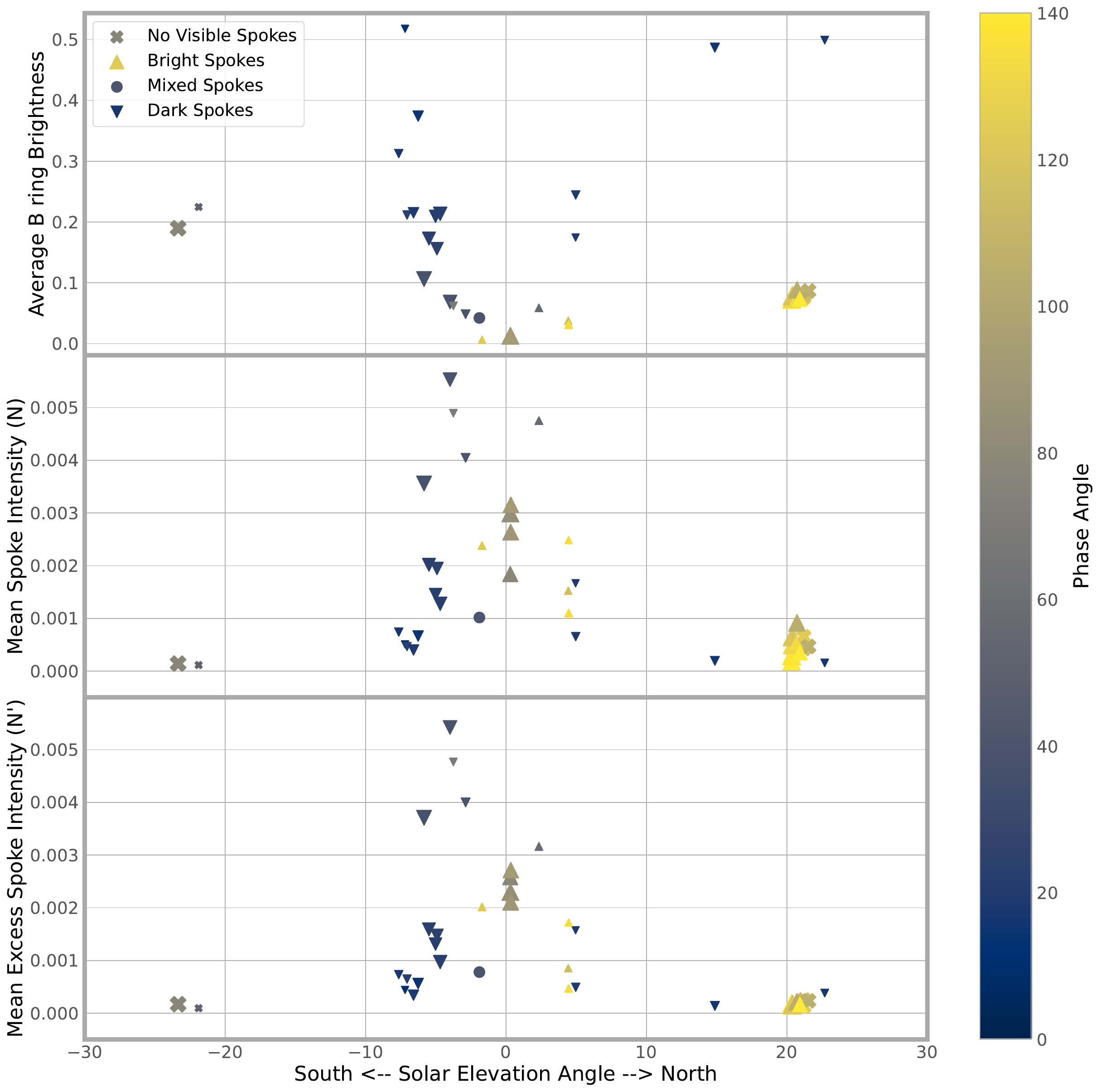}
    \caption{Timelines of average spoke intensity over the course of the Cassini mission derived from the profiles shown in Figures~\ref{fig:profs_avdat} and~\ref{fig:profs_avmin}. The shapes of the data points indicate the type of spokes visible in the images, and their sizes indicate the number of images considered in each observation, while the color indicates the observed phase angle. The top panel shows the average brightness of the B ring between 102,000 and 112,000 km. Note that the brightness {has a minimum at equinox, and declines rapidly between solar elevation angles of -10$^\circ$  and -5$^\circ$}. The bottom two panels show the mean normalized spoke signal $(\bar{n})$ and mean excess normalized spoke signal ($\bar{n}'$) between 102,000 km and 112,000 km above background levels, which show a clear rise and fall around equinox.}
    \label{fig:activity}
\end{figure*}

The profiles show in Figures~\ref{fig:profs_avdat} and~\ref{fig:profs_avmin}  show clear trends in spoke activity over the course of the Cassini mission. First of all, consider the observations taken at solar elevations below -2.5$^\circ$, which exclusively show dark spokes.  The two observations taken early in the Cassini Mission (when the solar elevation angle was below -20$^\circ$) both show flat profiles, consistent with the lack of spokes observed at this time. However, the observations obtained at solar elevation angles between -8$^\circ$ and $-6^\circ$ all show a subtle bump centered around 107,000 km, and this bump becomes more prominent for the observations obtained at solar elevation angles between $-6^\circ$ and $-2.5^\circ$. While the increase in the spoke signal is not monotonic between $-6^\circ$ and $-2.5^\circ$, it is still clear that the signal generally becomes stronger closer to equinox over this timespan. 

Moving on to the observations obtained around equinox, it is important to note that all the observations obtained at solar elevation angles between $-2^\circ$ and +2.5$^\circ$ either show mixed or bright spokes.  These spokes produce an obvious peak in the profiles obtained at solar elevation angles between $0^\circ$ and +2.5$^\circ$. However, the two observations obtained at solar elevation angles around -1.8$^\circ$ show a weaker spoke signal than observations taken slightly earlier and later. This is most likely related to the fact that one of these observations is the one showing mixed spokes (see Figure~\ref{fig: 108 mixed}). As mentioned above, mixed spokes probably arise in situations where slight changes in spoke properties can cause spokes to appear as bright or dark. More precisely, this means that $B(\alpha) \simeq I/F_b$ and so the factor of $|1-\frac{B(\alpha)}{I/F_b}|$ in  Equation~\ref{neq} is small, which means the same $rms$ variations in the spoke normal optical depth will yield a smaller value of $n(r)$. A similar phenomenon may also explain the lack of a strong signal in the second profile obtained around the same time. However, it is also worth noting that this observation was obtained at higher phase angles than the others from this time period, which could also be influencing the visibility of the spokes.  It is worth noting that the spokes in this particular observation appear to be more prevalent outside 112,000 km, which produces a peak that is displaced outwards compared to the other profiles (this can be seen more clearly in Figure~\ref{fig:profs_avmin}). Further work is therefore needed to ascertain whether this observation captured an unusual set of spokes, or if the specific lighting geometry highlighted spokes that are less prominent in other lit-side images.

Next, consider the observations at solar elevation angles between +4.5$^\circ$ and +5$^\circ$. Two of these observations show dark spokes, one of which has a bump around 107,000 km that is comparable to those found in profiles from observations with solar elevation angles around -5$^\circ$, while the other has a more subtle peak more similar to the profiles from solar elevation angles around -7$^\circ$.  The other three observations show bright spokes because they were obtained at higher phase angles. One of these profiles shows a bump between 105,000 and 110,000 km that is similar in size to the one showing dark spokes, while the others show less obvious signals. Remember that at these phase angles $B(\alpha) > I/F_b$, so the variations seen among these profiles may partially be due to variations in the observation geometry.

Lastly, we can note that all of the profiles obtained at solar elevation angles above 15$^\circ$ show no obvious bump in the outer B ring. While spokes could be seen in many of these observations, they were so rare that they do not produce very obvious features in the $n(r)$ profiles. However, it is worth noting that the bottommost profile in Figure~\ref{fig:profs_avmin} does show a very weak peak at 105,000 km, which corresponds to the location of the one spoke seen in this image sequence (see Figure~\ref{fig: final spoke1}). This indicates that these profiles can preserve some evidence of even weak spokes. Still, these profiles clearly demonstrate that the overall spoke activity at this time is low.

Another way to document these trends is to plot a measure of the overall strength of the spoke signal as a function of time or solar elevation angle. We therefore computed the average values of $\bar{n}(r)$ and $\bar{n}'(r)$ between 102,000 km and 112,000 km (which should contain most of the relevant spoke signals), and subtracted off  the average values of those same profiles between 99,000 km and 100,000 km and between 114,000 km and 115,000 km (which are largely spoke-free regions that give reasonable estimates of any background levels). We designate these quantities the mean spoke intensity $N$ and the mean excess spoke intensity $N'$, respectively. For reference, we also compute the average brightness of the background B ring between 102,000 km and 112,000 km. All three of these numbers are provided in Table~\ref{tab:profobs} and their values are plotted in Figure~\ref{fig:activity}. Note that we do not attempt to compute formal error bars on these quantities because their uncertainties are not primarily due to statistical noise but are instead primarily due to systematic variations that are difficult to quantify. {Instead, we can roughly estimate the typical fractional uncertainties on these parameters by considering groups of observations obtained at similar phase and solar elevation angles, which should have comparable spoke signals. In particular, the five observations obtained at solar elevation angles between -7.63$^\circ$ and -6.25$^\circ$ and phase angles between 11.3$^\circ$ and 22.7$^\circ$ yield $N$ and $N'$ values with fractional standard deviations of 0.26 and 0.29, respectively. Meanwhile, the eleven observations obtained at solar elevation angles between 20.34$^\circ$ and 21.55$^\circ$ and phase angles between 104.5$^\circ$ and 144$^\circ$ yield $N$ and $N'$ values with fractional standard deviations of 0.42 and 0.21, respectively. This indicates that the fractional uncertainties in these parameters are probably typically between 20\% and 50\%, although again it is important to recognize these uncertainties are dominated by systematic effects and so they cannot be treated like statistical error bars.}

The trends in $N$ and $N'$ shown in Figure~\ref{fig:activity} are generally consistent with one another, with $N'$ showing less dispersion at low activity levels and $N$ being substantially higher than $N'$ for several of the observations obtained after equinox that include a very small number of images. More importantly, both plots show that spokes are much more common and intense when the solar elevation angle is within 10$^\circ$ of zero. Indeed, there is a fairly clear trend of increasing spoke activity for solar elevation angles between  -8$^\circ$ and -4$^\circ$. \cite{McGhee05} previously noted a steep increase in the number and contrast of the spokes seen by HST at solar elevation angles between $-10^\circ$ and $-5^\circ$. These earlier data showed a strong correlation between emission and incidence angles that complicated efforts to interpret this trend, and so the new Cassini data can help clarify this situation. In particular, it now seems much less likely that this trend can be attributed entirely to changes in the lighting and viewing geometry.  Not only were most of the Cassini data during this time obtained at roughly the same emission angles, the brightness of the background B ring ($I/F_b$) actually declines by almost an order of magnitude. Since all of these observations include dark spokes, this means $B(\alpha) < I/F_b$ throughout this time period, so this decline in $I/F_b$ should also cause $|1-\frac{B(\alpha)}{I/F_b}|$ to decline. The observed rise in $N$ and $N'$ therefore requires an even larger increase in the $rms$ variations in the spoke normal optical depth \citep[provided the single-scattering approximation is valid for these observations, cf.][]{McGhee05}. These data  therefore indicate that spoke activity increases dramatically between solar elevation angles of -8$^\circ$ of $-4^\circ$. 

By contrast, the observed decline in $N$ and $N'$ between solar elevation angles of $-4^\circ$ and 0$^\circ$ is almost certainly primarily due to changes in the visibility of the spokes rather than trends in their typical optical depths. This time period is associated with the contrast reversal between dark and bright spokes, so the ratio of $B(\alpha)$ to $I/F_b$ crosses unity during this time, so the relationship between $N$ and the $rms$ variations in optical depth becomes more complicated. More careful photometric modeling will therefore need to be done in order to determine whether spoke activity continues to increase or levels off between solar elevation angles of $-4^\circ$ and 0$^\circ$.

Finally, while the data {are} far more sparse for positive solar elevation angles, the available data indicate that spoke activity around $+5^\circ$ is roughly comparable to the activity around $-5^\circ$. Hence there does not appear to be a strong asymmetry in the prevalence of spokes before and after equinox. In addition, we can note that the {single} observation of bright spokes at a solar elevation angle of $+2.35^\circ$ shows a comparable value of $N$ to that obtained from the dark-spoke observation around $-2.9^\circ$ (as mentioned above, the $N'$ value for this particular observation is probably underestimated because of the small number of images it contains). While it is not appropriate to directly compare the signals from dark and bright spokes due to them having very different values of $|1-\frac{B(\alpha)}{I/F_b}|$, this data point is at least broadly consistent with the prevalence of spokes being roughly symmetrical about equinox.

\section{Summary}
\label{summary}

The primary findings of this initial study can be summarized as follows:
\begin{itemize}
\item Cassini observed spokes over a wide range of lighting and viewing geometries.
\item Spokes on the unlit side of the rings can transition between being brighter or darker than the background ring due to changes in the illumination conditions.
\item Under appropriate conditions, different parts of a spoke can appear bright or dark due to spatial variations in the spoke particle properties, most likely trends in the spoke's  particle size distribution.
\item Spokes can be seen over the B ring even close to Saturn's solstice, but are rare and can sometimes only be clearly seen in extreme lighting geometries.
\item Spoke activity {is most intense} around equinox, when the solar elevation angle is within 10$^\circ$ of 0.
\end{itemize}

\pagebreak

\section*{Acknowledgements}

This work was supported by NASA Cassini Data Analysis Program Grant 80NSSC18K1397.  This work also made use of the Outer Planets Unified Search tool provided by the Planetary Data System Ring-Moon Systems Node. The authors would also like to thank A. Davies for helping with early stages of the data reduction pipeline, and the anonymous reviewers  for their helpful comments on this manuscript.

\bibliography{dustyrings}
\end{document}